\documentclass[10pt,conference]{IEEEtran}

\usepackage{xcolor}
\definecolor{code-blue}{RGB}{8,0,255}
\definecolor{code-red}{RGB}{255,83,112}
\definecolor{code-purple}{RGB}{102,0,153}
\usepackage[american]{babel}
\usepackage{microtype}

\usepackage{graphicx}  %Required
\usepackage{amsmath}
\usepackage{algorithm}
\usepackage{paralist}
\usepackage{framed}
\usepackage{amsfonts}
\usepackage[noend]{algorithmic}
\usepackage{multirow}
\usepackage{courier}
\usepackage{subcaption}
\usepackage[american]{babel}
\usepackage{microtype}

\ifCLASSINFOpdf
\else
\fi

\begin{document}
	%\title{Abstractive Code Summarization via Tree-Structured Advantage Actor-Critic Network Learning}
	\title{Improving Automatic Source Code Summarization via Deep Reinforcement Learning}
	\author{\IEEEauthorblockN{Yao Wan\textsuperscript{1}, Zhou Zhao\textsuperscript{1}, Min Yang\textsuperscript{2}, Guandong Xu\textsuperscript{3}, Haochao Ying\textsuperscript{1}, Jian Wu\textsuperscript{1}, Philip S. Yu\textsuperscript{4,5}}
		\IEEEauthorblockA{\textsuperscript{1}College of Computer Science and Technology, Zhejiang University, Hangzhou, China\\
			\textsuperscript{2}Shenzhen Institutes of Advanced Technology, Chinese Academy of Sciences, China\\
			\textsuperscript{3}Advanced Analytics Institute, University of Technology Sydney, Sydney, Australia\\
			\textsuperscript{4}Department of Computer Science, University of Illinois at Chicago, Illinois, USA\\
			\textsuperscript{5}Institute for Data Science, Tsinghua University, Beijing, China\\
			\{wanyao,zhaozhou,haochaoying,wujian2000\}@zju.edu.cn,\\ min.yang@siat.ac.cn, guandong.xu@uts.edu.au, psyu@uic.edu}
		}
	\maketitle
	
	\begin{abstract}
%		Code summarization provides a high level natural language description of the function performed by code, as it can benefit the software maintenance, code categorization and retrieval. To the best of our knowledge, most state-of-the-art approaches follow an encoder-decoder framework which encodes the code into a hidden space and then decode it into natural language space, suffering from two major drawbacks: a) their encoders only consider the sequential content of code, ignoring the tree structure which is also critical for the task of code summarization; b) their decoders are typically trained to predict the next word by maximizing the likelihood of next ground-truth word with previous ground-truth word given. However, it is expected to generate the entire sequence from scratch at test time. This discrepancy can cause an \textit{exposure bias} issue, making the learnt decoder suboptimal. In this paper, we incorporate an abstract syntax tree structure as well as sequential content of code snippets into a deep reinforcement learning framework (i.e., actor-critic network learning). The actor network provides the confidence of predicting the next word according to current state. On the other hand, the critic network evaluates the reward value of all possible extensions of the current state and can provide global and lookahead guidance for explorations. We employ an advantage reward composed of BLEU metric to train both networks. Comprehensive experiments on a real-world dataset show the effectiveness of our proposed model when compared with the state-of-the-art ones.
Code summarization provides a high level natural language description of the function performed by code, as it can benefit the software maintenance, code categorization and retrieval. To the best of our knowledge, most state-of-the-art approaches follow an encoder-decoder framework which encodes the code into a hidden space and then decode it into natural language space, suffering from two major drawbacks: a) Their encoders only consider the sequential content of code, ignoring the tree structure which is also critical for the task of code summarization; b) Their decoders are typically trained to predict the next word by maximizing the likelihood of next ground-truth word with previous ground-truth word given. However, it is expected to generate the entire sequence from scratch at test time. This discrepancy can cause an \textit{exposure bias} issue, making the learnt decoder suboptimal. In this paper, we incorporate an abstract syntax tree structure as well as sequential content of code snippets into a deep reinforcement learning framework (i.e., actor-critic network). The actor network provides the confidence of predicting the next word according to current state. On the other hand, the critic network evaluates the reward value of all possible extensions of the current state and can provide global guidance for explorations. We employ an advantage reward composed of BLEU metric to train both networks. Comprehensive experiments on a real-world dataset show the effectiveness of our proposed model when compared with some state-of-the-art methods.
		%Code summarization provides a high level description of the function performed by code, as it can benefit the software maintenance, code categorization and retrieval. 
		%%In order to provide high-quality summarization, many existing 
		%Most state-of-the-art approaches follow an encoder-decoder framework, which encodes the code sequence into a hidden state, and generates word by maximizing the likelihood of next ground-truth word with previous ground-truth word given. These approaches are limited in two-fold: a) on encoding, they only consider the sequential content of code, ignoring the tree structure which is also critical for the task of code summarization; b) on decoding, they are teacher-forcing based and suffer from the exposure bias so that the learnt decoder is suboptimal. In this paper, we incorporate an abstract syntax tree structure as well as sequential content of code snippets into an actor-critic network learning framework. The actor network provides the confidence of predicting the next word according to current state. On the other hand, the critic network evaluates the reward value of all possible extentions of the current state and can provide global and lookahead guidance for explorations. We employ an advantage reward composed of BLEU metric to train both networks. Comprehensive experiments on a real-world dataset show the effectiveness of our proposed model when compared with the state-of-the-art ones.
		
	\end{abstract}
	

	% \fi
	\IEEEpeerreviewmaketitle
	
	\section{Introduction}
	In the life cycle of software development (e.g., implementation, testing and maintenance), nearly 90\% of effort is used for maintenance, and much of this effort is spent on understanding the maintenance task and related software source codes \cite{lientz1980software}. Thus, documentation which provides a high level description of the task performed by code is always a must for software maintenance. Even though various techniques have been developed to facilitate the programmer during the implementation and testing of software, documenting code with comments remains a labour-intensive task, making few real-world software projects adequately document the code to reduce future maintenance costs \cite{de2005study, kajko2005survey}. 
	\begin{figure*}[!t]
		\centering
		\begin{subfigure}[b]{3.2in}
			\includegraphics[width=\textwidth]{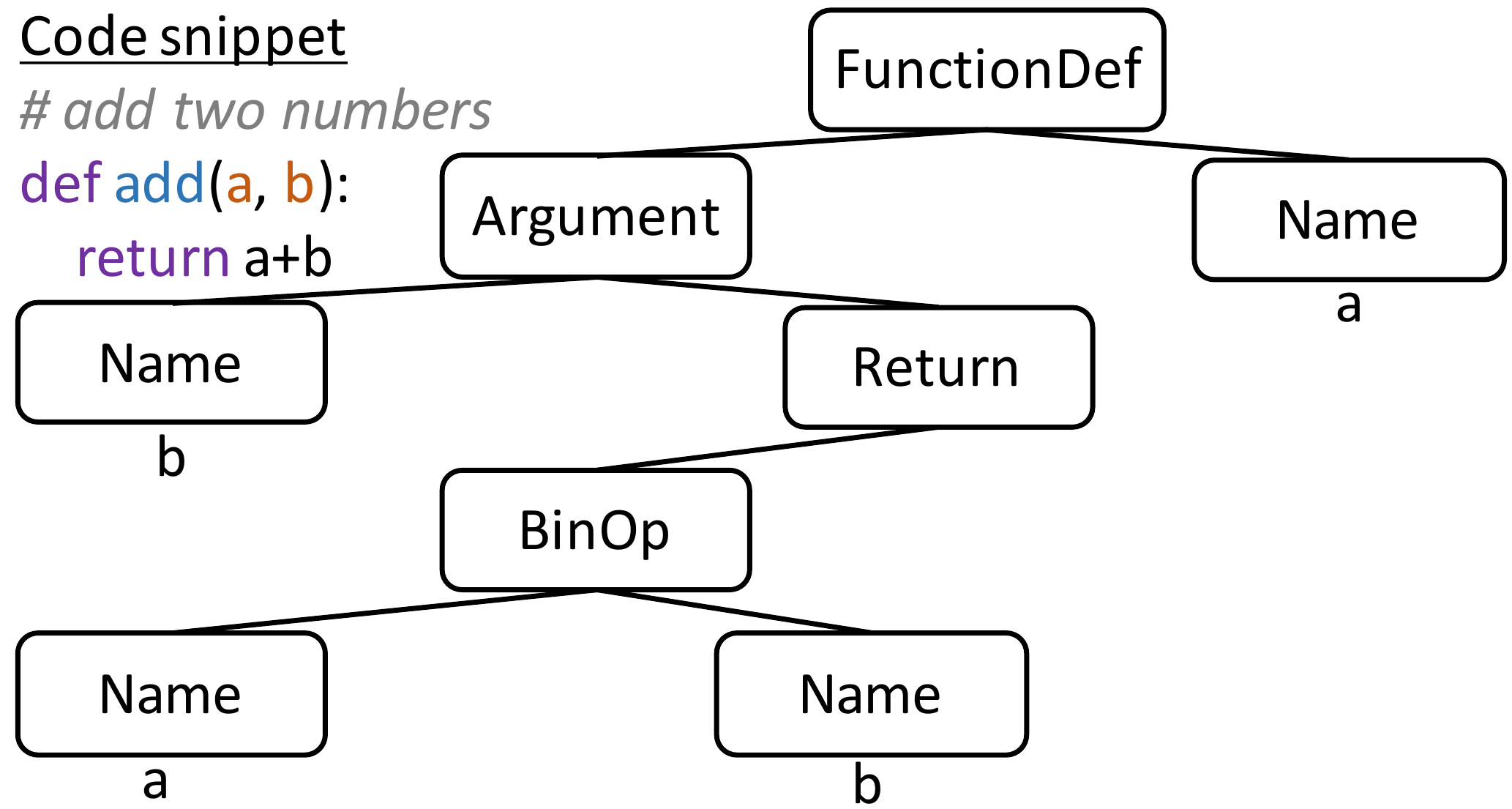}
			\caption{An example of abstractive syntax tree (AST).}
			\label{fig_ast}
		\end{subfigure}
		\begin{subfigure}[b]{3.2in}
			\includegraphics[width=\textwidth]{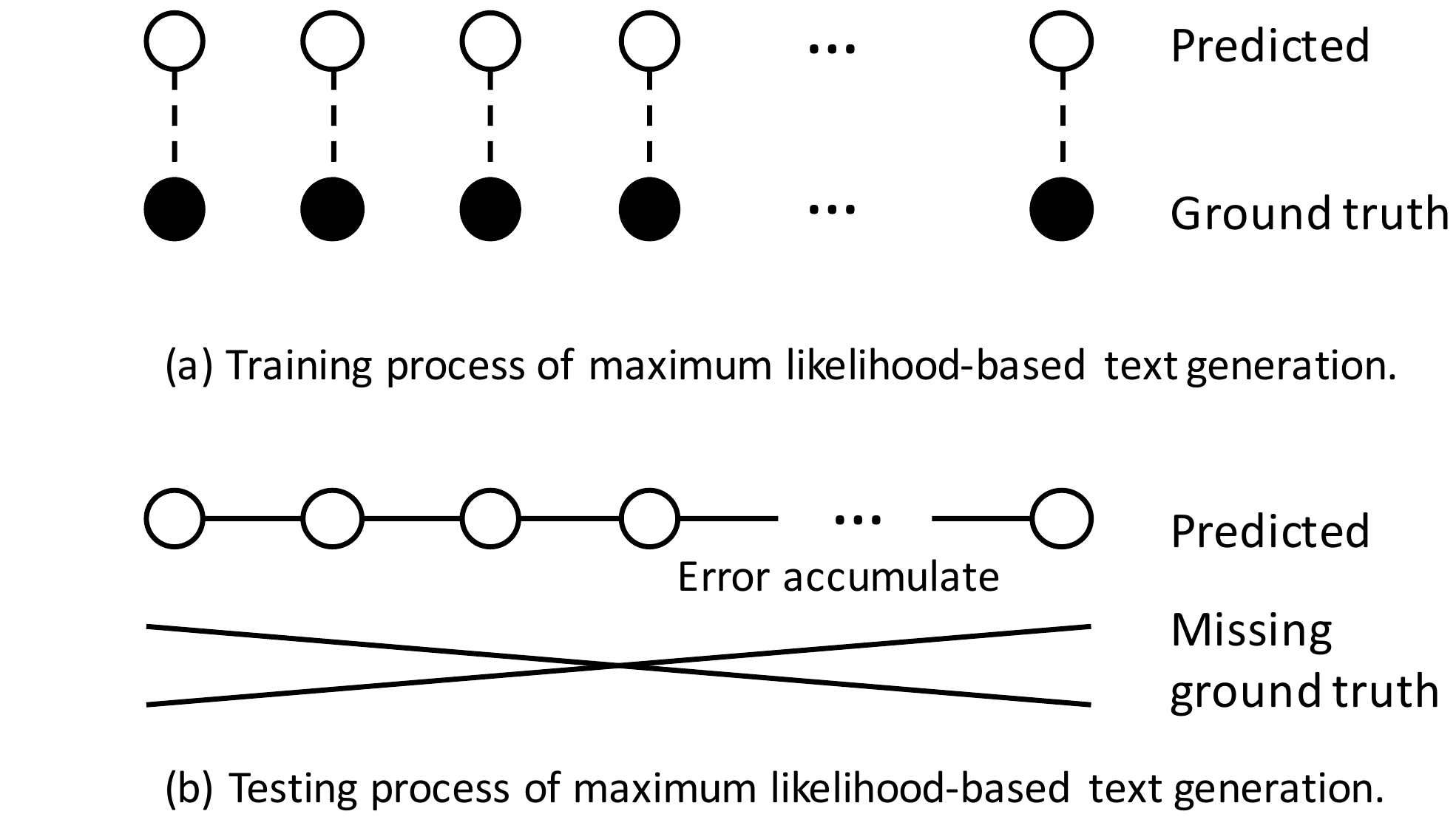}
			\caption{The limitation of maximum likelihood based text generation. }
			\label{fig_exposure_bias}
		\end{subfigure}
		\caption{An illustration of the motivation of our paper. Traditional methods suffer from the following two limitations: a) On representing the code, the structure information of code is always ignored. b) Traditional maximum likelihood based methods suffer from the exposure bias issue.}
		\label{fig_motivation}
	\end{figure*}
	It's nontrivial for a novice programmer to write good comments for source codes. A good comment should at least has the following characteristics: a) Correctness. The comments should correctly clarify the intent of code. b) Fluency. The comments should be fluent natural languages that can be easily read and understood by maintainers. c) Consistency. The comments should follow a standard style/format for better code reading. Code summarization is a task that tries to comprehend code and automatically generate descriptions directly from the source code. The summarization of code can also be viewed as a form of document expansion. Successful code summarization can not only benefit the maintenance of source codes \cite{movshovitz2013natural,iyer2016summarizing}, but also be used to improve the performance of code search using natural language queries \cite{yang2016query,nie2016query} and code categorization \cite{nguyen2017automatic}. 
	
	\noindent\textbf{Motivation.} Recent research has made inroads towards automatic generation of natural language descriptions of software. As far as we know, most of existing code summarization methods learn the semantic representation of source codes based on statistical language models \cite{oda2015learning,movshovitz2013natural}, and then generate comments based on templates or rules \cite{sridhara2010towards}. With the development of deep learning, some neural translation models \cite{allamanis2016convolutional,iyer2016summarizing,haije2016automatic} have also been introduced for code summarization, which mainly follow an encoder-decoder framework. They generally employ recurrent neural networks (RNN e.g., LSTM \cite{hochreiter1997long}) to encode the code snippets and utilize another RNN to decode that hidden state to coherent sentences. These models are typically trained to maximize the likelihood of the next word on the assumption that previous words and ground-truth are given. These models are limited from two aspects: a) The code sequential and structural information is not fully utilized on feature representation, which is critical for code understanding. For example, given two simple expressions ``\texttt{f=a+b}" and ``\texttt{f=c+d}", although they are quite different as two lexical sequences, they share the same structure (e.g., abstractive syntax tree). 
	b) These models, also termed ``teacher-forcing", suffer from the \textit{exposure bias} since in testing time the ground-truth is missing and previously generated words from the trained model distribution are used to predict the next word \cite{ranzato2015sequence}. Figure \ref{fig_motivation}(b) presents a simple illustration of the discrepancy among training and testing process in these classical encoder-decoder models. In the testing phase, this exposure bias makes error accumulated and makes these models suboptimal, not able to generate those words which are appropriate but with low probability to be drawn in the training phase.
	
	\noindent\textbf{Contribution. }In this paper, we aim to address these two mentioned issues. To effectively capture the structural (or syntactic) information of code snippets, we employ abstract syntax tree (AST) \cite{baxter1998clone}, a data structure widely used in compilers, to represent the structure of program code. Figure \ref{fig_ast} shows an example of Python code snippet and its corresponding AST. The root node is a composite node of type \texttt{FunctionDef}, while the leaf nodes which are typed as \texttt{Name} are tokens of code snippets. It's worth mentioning that the tokens from AST parsing may be different from those from word segmentation. In our paper, we consider both of them. 
	We parse the code snippets into ASTs, and then propose an AST-based LSTM model \cite{tai2015improved} to represent the structure of code. We also use another LSTM model \cite{hochreiter1997long} to represent the sequential information of code. Besides, we apply a hybrid attention layer to fuse the structure representation and sequential representation of code on predicting the word, considering the alignment between predicted word and source word.
	
	To overcome the exposure bias, we draw on the insights of deep reinforcement learning, which integrates exploration and exploitation into a whole framework. 
	Instead of learning a sequential recurrent model to greedily look for the next correct word, we utilize an actor network and a critic network to jointly determine the next best word at each time step. The actor network, which provides the confidence of predicting the next word according to current state, serves as a local guidance. The critic network, which evaluates the reward value of all possible extensions of the current state, serves as a global guidance. Our framework is able to include the good words that are with low probability to be drawn by using the actor network alone. To learn these two networks more efficiently, we start with pretraining an actor network using standard supervised learning with cross entropy loss, and pretraining a critic network with mean square loss. Then, we update the actor and critic networks according to the advantage reward composed of BLEU metric via policy gradient. We summarize our main contributions as follows.
	\begin{itemize}
		\item We propose a more comprehensive representation method for source code, with one AST-based LSTM for the structure of source code, and another LSTM for the sequential content of source code. Furthermore, a hybrid attention layer is applied to fuse these two representations.
		\item To the best of our knowledge, it is the first time that we propose an advanced deep reinforcement learning framework, named actor-critic network, to cope with the exposure bias issue existing in most traditional maximum likelihood-based code summarization frameworks.
		\item We validate our proposed model on a real-world dataset of 108,726 Python code snippets. Comprehensive experiments show the effectiveness of our proposed model when compared with some state-of-the-art methods. %To facilitate other researchers to repeat our experiments, we release the dataset and source code used in this paper in \textit{https://github.com/wanyao1992/code\_summarization\_public}.
	\end{itemize}
	\textbf{Organization.} The remainder of this paper is organized as follows. We provide some background knowledge on neural language model, RNN encoder-decoder model and reinforcement learning in Section \ref{sec_background} for a better understanding of our proposed model. We also formally define the problem in Section \ref{sec_background}. Section \ref{sec_framework} gives an overview of our proposed framework. Section \ref{sec_hybrid_emb} presents a hybrid embedding approach for code representation. Section \ref{sec_drl_model} shows our proposed deep reinforcement learning framework (i.e., actor-critic network). Section \ref{sec_experiments} describes the dataset used in our experiment and shows the experimental results and analysis. Section \ref{sec_threats} shows some threats to validity and limitations existing in our model. Section \ref{sec_relatedwork} highlights some works related to this paper. Finally, we conclude this paper in Section \ref{sec_conclusion}. %and give some future research directions
	
	\section{Background}\label{sec_background}
	As we declared before, the code summarization task can be seen as a text generation task given the source code. In this section, we first present some background knowledge on text generation which will be used in this paper, including language model, attentional RNN encoder-decoder model and reinforcement learning for better decoding.
	To start with, we introduce some basic notations and terminologies. 
	Let $\mathbf{x}=(x_1, x_2,\ldots, x_{|\mathbf{x}|})$ denote a sequence of source code snippet, $\mathbf{y}=(y_1, y_2,\ldots, y_{|\mathbf{y}|})$ denote a sequence of generated words, where {\scriptsize $|\cdot|$} denotes the length of sequence. Let $T$ denote the maximum step of decoding in the encoder-decoder framework. We will often use notation $\mathbf{y}_{m \ldots l}$ to refer to subsequences of the form $(y_m, \ldots , y_l)$. $\mathcal{D}=\{(\mathbf{x}_1,\mathbf{y}_1), (\mathbf{x}_2,\mathbf{y}_2),\ldots, (\mathbf{x}_N,\mathbf{y}_N)\}$ is the training dataset, where $N$ is the size of training set.
	
	\subsection{Language Model}
	%Deep Learning Code Fragments for Code Clone Detection
	%
	Language model computes the probability of occurrence of a number of words in a particular sequence. The probability of a sequence of $T$ words $\{y_1 , \ldots, y_T \}$ is denoted as $p(y_1 , \ldots, y_T)$. Since the number of words coming before a word, $y_i$ , varies depending on its location in the input document, $p(y_1 , \ldots, y_T)$ is usually conditioned on a window of $n$ previous words rather than all previous words:
	\begin{equation}
	p(y_{1:T})=\prod_{i=1}^{i=T}p(y_i|y_{1:i-1})\approx \prod_{i=1}^{i=T}p(y_i|y_{i-(n-1):i-1}).
	\end{equation}
	
	This kind of n-grams approach suffers apparent limitations \cite{rosenfeld2000two,mnih2012fast}. For example, the n-gram model probabilities can not be derived directly from the frequency counts, because models derived this way have severe problems when confronted with some n-grams that have not been explicitly seen before. 
	
\begin{figure}[!t]
		\centering
		\includegraphics[width=0.45\textwidth]{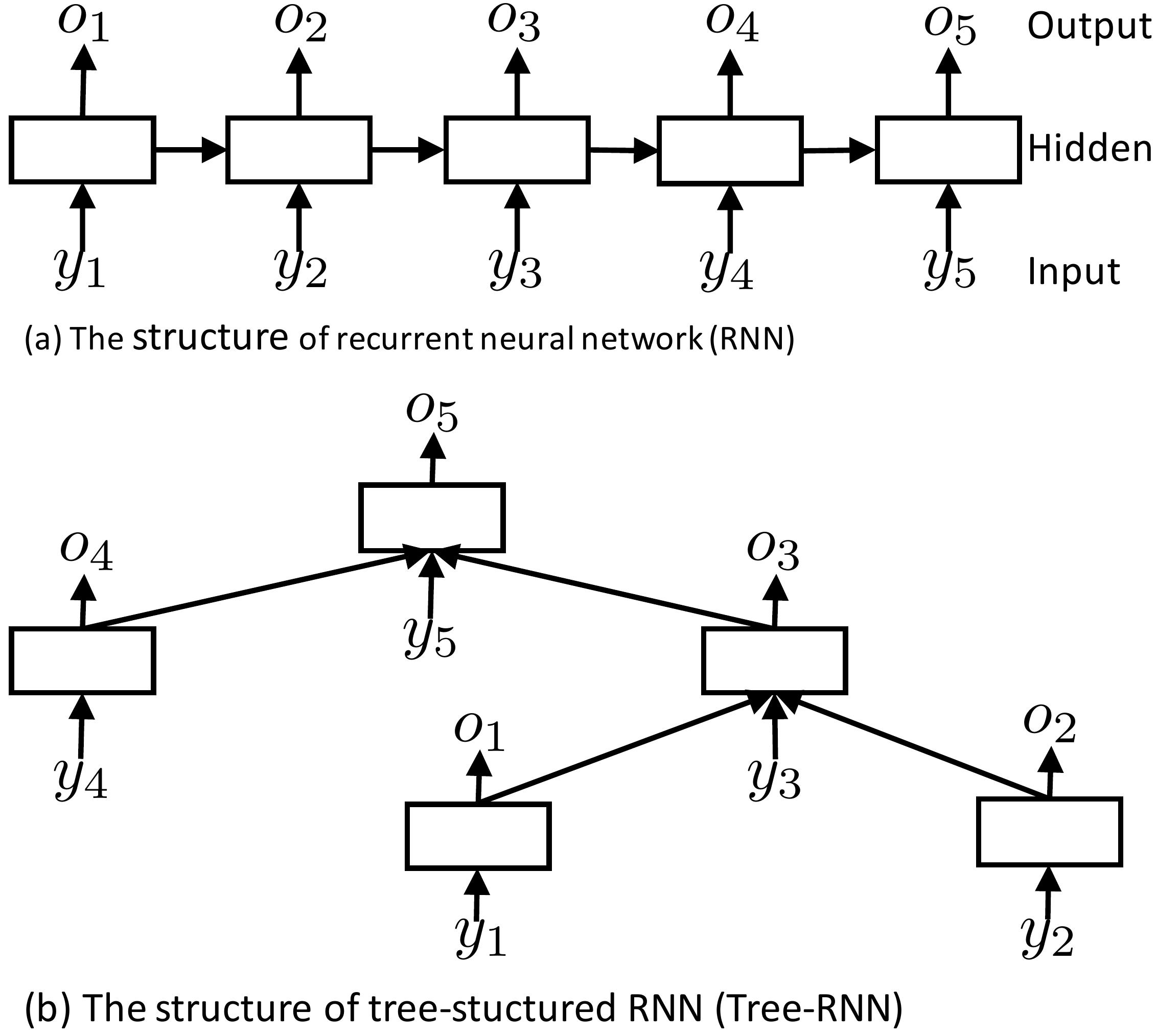}
		\caption{RNN and Tree-RNN (adapted from \cite{tai2015improved}).}
		\label{fig_lstm_treelstm}
	\end{figure}

	The neural language model is a language model based on neural networks. Unlike the n-gram model which predicts a word based on a fixed number of predecessor words, a neural language model can predict a word by predecessor words with longer distances. Figure \ref{fig_lstm_treelstm}(a) shows the basic structure of a RNN. The neural network includes three layers, that is, an input layer which maps each word to a vector, a recurrent hidden layer which recurrently computes and updates a hidden state after reading each word, and an output layer which estimates the probabilities of the following word given the current hidden state. The RNN reads the words in the sentence one by one, and predicts the possible following word at each time step. At step $t$, it estimates the probability of the following word $p(y_{t+1}|y_{1:t})$ by the following steps: First, the current word $y_t$ is mapped to a vector by the input layer $e$. Then, it generates the hidden state  $\mathbf{h}_t$ at time $t$ according to the previous hidden state $\mathbf{h}_{t-1}$ and the current input $y_t$:
	\begin{equation}
	\mathbf{h}_t = f(\mathbf{h}_{t-1}, e(y_t)).
	\end{equation}
	
	Here, two common options for $f$ are long short-term memory (LSTM) \cite{hochreiter1997long} and the gated recurrent unit (GRU) \cite{li2015gated}. Finally, the $p(y_{t+1}|y_{1:t})$ is predicted according to the current hidden state $\mathbf{h}_t$ :
	\begin{equation}
	p(y_{t+1}|y_{1:t}) = g(\mathbf{h}_t),
	\end{equation}
	where $g$ is a stochastic output layer (typically a softmax for discrete outputs) that generates output tokens.
	
	\subsection{Attentional RNN Encoder-Decoder Model}
	RNN encoder-decoder has two recurrent neural networks. 
	The encoder transforms the code snippet $\mathbf{x}$ into a sequence of hidden states $(\mathbf{h}_1 ,\mathbf{h}_2, \ldots, \mathbf{h}_{|\mathbf{x}|})$ with a RNN, while the decoder uses another RNN to generate one word $y_{t+1}$ at a time in the target space. 
	
	\subsubsection{Encoder}
	As a RNN, the encoder has a hidden state, which is a fixed-length vector. At the time step $t$, the encoder computes the hidden state $\mathbf{h}_t$ by:
	\begin{equation}
	\mathbf{h}_t = f(\mathbf{h}_{t-1}, \mathbf{c}_{t-1}, e(\mathbf{x}_t))).
	\end{equation}
	
	Here, $f$ is the hidden layer which has two main options, i.e., LSTM and GRU. The last symbol of $\mathbf{x}$ should be an end-of-sequence ($<eos>$) symbol which notifies the encoder to stop and output the final hidden state $\mathbf{h}_T$, which is used as a vector representation of $\mathbf{x}$.
	
	\subsubsection{Decoder}
	The output of the decoder is the target sequence $\mathbf{y}=(y_1 , \cdots, y_T)$. One input of the decoder is a $<start>$ symbol denoting the beginning of the target sequence. At the time step $t$, the decoder computes the hidden state $\mathbf{h}_t$ and the conditional distribution of the next symbol $y_{t+1}$ by:
	\begin{equation}
	p(y_{t+1}|y_t) = g(\mathbf{h}_{t}, \mathbf{c}_{t}),
	\end{equation}
	where $g$ is a stochastic output layer and $\mathbf{c}_{t}$ is the distinct context vector for $y_t$, computed by:
	\begin{equation}
	\mathbf{c}_t=\sum_{j=1}^{|\mathbf{x}|}\alpha_{t,j}\mathbf{h}_j, 
	\end{equation}
	where $\alpha_{t,j}$ is the attention weight of $y_t$ on $\mathbf{h}_j$ \cite{bahdanau2014neural}.
	
	\subsubsection{Training Goal}
	The encoder and decoder networks are jointly trained to maximize the following objective:
	
	\begin{equation}\label{eq_actor_objective}
	\underset{\theta}{\max}\mathcal{L}(\theta) = \underset{\theta}{\max}\underset{(\mathbf{x},\mathbf{y})\sim\mathcal{D}}{\mathbb{E}}\log p(\mathbf{y}|\mathbf{x};\theta),
	\end{equation}
	where $\theta$ is the set of the model parameters. We can see that this classical encoder-decoder framework targets on maximizing the likelihood of ground-truth word conditioned on previously generated words. As we have mentioned above, the maximum likelihood based encoder-decoder framework suffers the exposure bias issue. Motivated by this, we introduce the reinforcement learning technique for better decoding.  
	
	\subsection{Reinforcement Learning for Better Decoding}
	The reinforcement learning is an approach that interacts with the real environment and learns the optimal policy from the reward signal. It tries to generate text from scratch without ground truth in the testing phase. Under this approach, the text generation process can be viewed as a Markov Decision Process (MDP) $\{S, A, P, R, \gamma\}$. In the MDP setting, state $\mathbf{s}_t$ at time step $t$ consists of the source code snippets $\mathbf{x}$ and the words/actions predicted until $t$, ${y_0,y_1,\ldots,y_t}$. The action space is the dictionary $\mathcal{Y}$ that the words are drawn from, \textit{i.e.,} $y_t \subset \mathcal{Y}$. With the definition of the state, the state transition function $P$ is $\mathbf{s}_{t+1} = \{\mathbf{s}_t, y_{t+1}\}$, where the action $y_{t+1}$ becomes a part of the next state $\mathbf{s}_{t+1}$ and the reward $r_{t+1}$ is received. $\gamma \in [0,1]$ is the discount factor.
	The objective of generation process is to find a policy that maximizes the expected reward of generation sentence sampled from the model's policy:
	
	\begin{equation}
	\underset{\theta}{\max}\mathcal{L}(\theta) = \underset{\theta}{\max}\mathbb{E}_{\underset{\hat{\mathbf{y}}\sim P_{\theta}(\cdot|\mathbf{x})}{\mathbf{x}\sim \mathcal{D}}}[R(\hat{\mathbf{y}},\mathbf{x})],
	\end{equation}
	where $\theta$ is the parameter of policy needed to be learnt, $\mathcal{D}$ is the training set, $\hat{\mathbf{y}}$ is the predicted actions/words,  and $R$ is the reward function. Our problem can be formulated as follows. 
	\vspace{1em}
	
	\noindent\fbox {\begin{minipage}{0.46\textwidth}
			Given a code snippet $\mathbf{x}=(x_1, x_2,\ldots, x_{|\mathbf{x}|})$, our goal is to find a policy that generates a sequence of words $\mathbf{y}=(y_1, y_2,\ldots, y_{|\mathbf{y}|})$ from dictionary $\mathcal{Y}$ with the objective of maximizing the expected reward. 
		\end{minipage}
	}
	\vspace{1em}
	
	To learn the policy, many approaches have been proposed, which are mainly categorized into two classes \cite{sutton1998introduction}. a) The policy-based approaches (e.g., REINFORCE \cite{williams1992simple}) which optimizes the policy directly via policy gradient.  b) The value-based approaches (e.g., Q-learning \cite{watkins1992q}) which learns the Q-function, and in each time the agent selects the action with highest Q-value. It has been verified that the policy-based methods may suffer from a variance issue and the value-based methods may suffer from a bias issue \cite{keneshloo2018deep}. Thus in our paper, we adopt the actor-critic learning method which is a more advanced technique that has the advantage of both policy- and value-based methods. %To have a deeper understanding on reinforcement learning, we put the detailed derivation of policy gradient in Appendix A, which will be used in our model training process.
	
	\section{Overview of Proposed Framework}\label{sec_framework}
\begin{figure}[!t]
	\centering
	\includegraphics[width=0.48\textwidth]{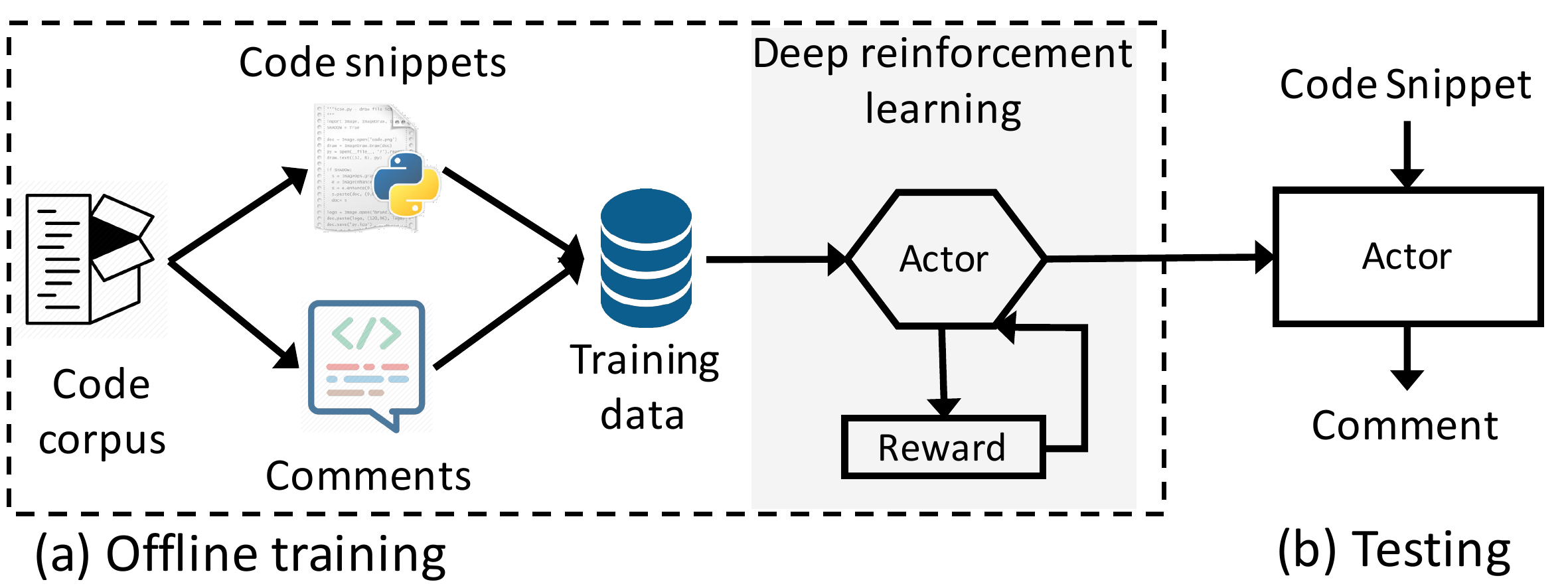}
	\caption{An overall workflow of getting a trained model.}
	\label{fig_workflow}
	%		\vspace{-1em}
\end{figure}
\begin{figure*}[!t]
	\centering
	\includegraphics[width=0.96\textwidth]{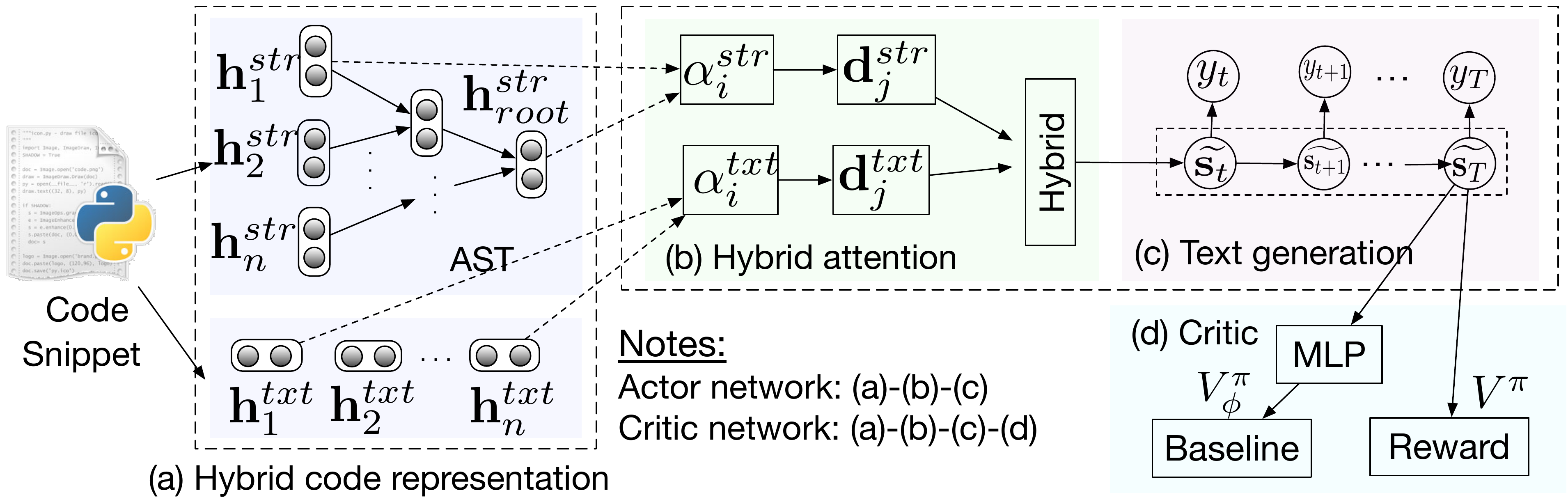}
	\caption{An overview of our proposed deep reinforcement learning framework for code summarization.}
	\label{fig_framework}
\end{figure*}
In this section, we firstly have a simple overview on the workflow of how to get a trained model for code summarization. Then we present an overview of the network architecture of our proposed deep reinforcement learning based model. Figure \ref{fig_workflow} shows the overall workflow of how to get a trained model. It includes an offline training stage and an online summarization stage. In the training stage, we prepare a large-scale corpus of annotated $<code, comment>$ pairs. The annotated pairs are then fed into our proposed deep reinforcement learning model for training. After training, we can get a trained actor network. Then, given a code snippet, corresponding comment can be generated by the trained actor network. 

Figure \ref{fig_framework} is an overview of the network architecture of our proposed deep reinforcement learning based model. The architecture of our model follows the actor-critic framework \cite{konda2000actor}, which has been successfully adopted in the decision-making scenarios such as AlphaGo \cite{silver2016mastering}. We split the framework into four submodules. (a) Hybrid code representation (cf. Sec. \ref{sec_hybrid_emb}). This module is used to represent the source code into a hidden space, which is also called encoder in the encoder-decoder framework. (b) Hybrid attention (cf. Sec. \ref{sec_hybrid_attention}). On decoding the encoded hidden space into the comment space, the attention layer is used to assign different weights to the code snippet tokens for better generation. (c) Text generation (cf. Sec. \ref{sec_text_generation}). This module is a RNN-based generative network, which is used to generate the next word based on previous generated words. (d) Critic (cf. Sec. \ref{sec_critic}). This module is used to evaluate whether the generated word is good or not.

Since the generated tokens on (d) can also been seen as actions, we can also called the process (a)-(b)-(c) as actor network. Compared with the architecture of traditional encoder-decoder framework, our proposed model has an additional critic module used to evaluate the value of action taken under current state. The process (a)-(b)-(c)-(d) can also be called as critic network. We can see that the actor and critic networks share the modules (a)-(b)-(c), reducing the number of learning parameters a lot. We will elaborate each component in this framework in the following sections.

\section{Hybrid Representation of Code}\label{sec_hybrid_emb}%A Hybrid Embedding Approach for Code Representation
Different from previous methods that just utilize sequential tokens to represent code, we also consider the structure information of source code. In this section, we present a hybrid embedding approach for code representation. We apply an LSTM to represent the lexical level of code, and an AST-based LSTM to represent the syntactic level of code. %We will elaborate our proposed actor-critic network in the next section.
\subsection{Lexical Level}
The key insight into lexical level representation of source code is that comments are always extracted from the lexical of code, such as the function name, variable name and so on. It's apparent that we apply an RNN (e.g., LSTM) to represent the sequential information of source code. In our paper, the LSTM is adopted.

\subsection{Syntactic Level}
In executing a program, a compiler decomposes a program into constituents and produces intermediate code according to the syntax of the language. 
AST is one type of intermediate code that represents the hierarchical syntactic structure of a program \cite{aho1986compilers}. We represent the syntactic level of source code from the aspect of AST embedding.
Similar to a traditional LSTM unit, we propose AST-based LSTM where the LSTM unit also contains an input gate, a memory cell and an output gate. Different from a standard LSTM unit which only has one forget gate for its previous unit, an AST-based LSTM unit contains multiple forget gates. 
Given an AST, for any node $j$, let the hidden state and memory cell of its $l$-th child be $h_{jl}$ and $c_{jl}$ respectively. Refer to \cite{tai2015improved}, the hidden state is updated as follows:
\begin{eqnarray}\label{eq_lstm}
%	\widetilde{\mathbf{h}_j}&=&\sum_{k\in C(j)}\mathbf{h}_k, \nonumber \\
\mathbf{i}_j&=&\sigma(\mathbf{W}^{(i)}\mathbf{x}_j+\sum_{l=1}^{N}\mathbf{U}_{l}^{(i)}\mathbf{h}_{jl}+\mathbf{b}^{(i)}), \nonumber \\
\mathbf{f}_{jk}&=&\sigma(\mathbf{W}^{(f)}\mathbf{x}_j+\sum_{l=1}^{N}\mathbf{U}_{kl}^{(f)}\mathbf{h}_{jl}+\mathbf{b}^{(f)}), \nonumber \\
\mathbf{o}_j&=&\sigma(\mathbf{W}^{(o)}\mathbf{x}_j+\sum_{l=1}^{N}\mathbf{U}_{l}^{(o)}\mathbf{h}_{jl}+\mathbf{b}^{(o)}), \nonumber \\
\mathbf{u}_j&=&\tanh(\mathbf{W}^{(u)}\mathbf{x}_j+\sum_{l=1}^{N}\mathbf{U}_{l}^{(u)}\mathbf{h}_{jl}+\mathbf{b}^{(u)}), \nonumber \\
\mathbf{c}_j&=&\mathbf{i}_j\odot \mathbf{u}_j + \sum_{l=1}^{N}\mathbf{f}_{jl}\odot \mathbf{c}_{jl}, \nonumber\\
\mathbf{h}_j &=& \mathbf{o}_j\odot \tanh(\mathbf{c}_j),
\end{eqnarray}
where $k=1,2,\cdots,N$. Each of $\mathbf{i}_j$, $\mathbf{f}_{jk}$, $\mathbf{o}_j$ and $\mathbf{u}_j$ denotes an input gate, a forget gate, an output gate, and a state for updating the memory cell, respectively. $\mathbf{W}^{(\cdot)}$ and $\mathbf{U}^{(\cdot)}$ are weight matrices, $\mathbf{b}^{(\cdot)}$ is a bias vector, and $\mathbf{x}_j$ is the word embedding of the $j\text{-}$th node. $\sigma(\cdot)$ is the logistic function, and the operator $\odot$ denotes element-wise multiplication between vectors. It's worth mentioning that when the tree is simply a chain, namely $N=1$, the AST-based LSTM unit reduces to the standard LSTM. Figure \ref{fig_lstm_treelstm} shows the structure of RNN and Tree-RNN.

Notice that the number of children $N$ varies for different nodes of different ASTs, which may cause problem in parameter-sharing. For simplification, we transform the generated ASTs to binary trees by the following two steps which have been adopted in \cite{weisupervised}: a) Split nodes with more than 2 children, generate a new right child together with the old left child as its children, and then put all children except the leftmost as the children of this new node. Repeat this operation in a top-down way until only nodes with 0, 1, 2 children left; b) Combine nodes with 1 child with its child.
\section{Deep Reinforcement Learning for Code Summarization}\label{sec_drl_model}
In this section, we introduce the advanced deep learning framework named actor-critic network, which has been successfully used in the AlphaGo \cite{silver2016mastering}. We introduce the actor and critic network respectively and then present how to train them simultaneously.

\subsection{Actor Network}
After obtaining the representation of code snippet, we need to decode it into comment. Here we describe how we generate comment from the hidden space with a hybrid attention layer.
\subsubsection{Hybrid Attention} \label{sec_hybrid_attention}
Different parts of the code make different contributions to the final output of comment. We adopt an attention mechanism \cite{bahdanau2014neural} which has been successfully used in neural machine translation. In the attention layer, we have two attention scores, one $\mathbf{\alpha}_t^{str}(j)$ for structural representation and another $\mathbf{\alpha}_t^{txt}(j)$ for sequential representation of code. At $t\text{-}$th step of the decoding process, the attention scores $\mathbf{\alpha}_t^{str}(j)$ and $\mathbf{\alpha}_t^{txt}(j)$ are calculated as follows: 

\begin{equation}
\mathbf{\alpha}_t^{str}(j)=\frac{\exp(\mathbf{h}_j^{str}\cdot \mathbf{s}_t)}{\sum_{k=1}^{n}\exp(\mathbf{h}_k^{str} \cdot \mathbf{s}_t)},\\  \mathbf{\alpha}_t^{txt}(j)=\frac{\exp(\mathbf{h}_j^{txt}\cdot \mathbf{s}_t)}{\sum_{k=1}^{n}\exp(\mathbf{h}_k^{txt} \cdot \mathbf{s}_t)},
\end{equation}
where $n$ is the number of code tokens; $\mathbf{h}_j^{(\cdot)}\cdot \mathbf{s}_t$ is the inner project of $\mathbf{h}_j^{(\cdot)}$ and $\mathbf{s}_t$, which is used to directly calculate the similarity score between $\mathbf{h}_j^{(\cdot)}$ and $\mathbf{s}_t$. The $t\text{-}$th context vector $\mathbf{d}_t^{(\cdot)}$ is calculated as the summarization vector weighted by $\mathbf{\alpha}_t^{(\cdot)}(j)$:

\begin{equation}
\mathbf{d}_t^{str} = \sum_{t=1}^{n}\mathbf{\alpha}_t^{str}(j)\mathbf{h}_j^{str},\ \  \mathbf{d}_t^{txt} = \sum_{t=1}^{n}\mathbf{\alpha}_t^{txt}(j)\mathbf{h}_j^{txt}.
\end{equation}

To integrate the structural context vector and the textual vector, we concatenate them firstly and then feed them into an one-layer linear network:

\begin{equation}
\mathbf{d}_t = \mathbf{W}_{\mathbf{d}_t}[\mathbf{d}_t^{str}; \mathbf{d}_t^{txt}] + \mathbf{b}_{\mathbf{d}_t}),
\end{equation}
where $[\mathbf{d}_t^{str}; \mathbf{d}_t^{txt}]$ is the concatenation of $\mathbf{d}_t^{str}
$ and $\mathbf{d}_t^{txt}$.
%To incorporate the attention mechanism into the actor network,
The context vector is then used for the $(t+1)\text{-}$th word prediction by putting an additional hidden layer $\widetilde{\mathbf{s}_t}$:

\begin{equation}
\widetilde{\mathbf{s}_t} = \tanh(\mathbf{W}_c[\mathbf{s}_t;\mathbf{d}_t]+\mathbf{b}_d),
\end{equation}
where $[\mathbf{s}_t;\mathbf{d}_t]$ is the concatenation of $\mathbf{s}_t$ and $\mathbf{d}_t$. 

\subsubsection{Text Generation}\label{sec_text_generation}
The model predicts the $t\text{-}$th word by using a softmax function. Let $p_{\pi}$ denote a policy $\pi$ determined by the actor network, $p_{\pi}(y_t|\mathbf{s}_t)$ denote the probability distribution of generating $t\text{-}$th word $y_t$, we can get the following equation:

\begin{equation}
p_{\pi}(y_t|\mathbf{s}_t)=softmax(\mathbf{W}_s\widetilde{\mathbf{s}_t}+\mathbf{b}_s).
\end{equation}

\subsection{Critic Network}\label{sec_critic}
Unlike traditional encoder-decoder framework that generates sequence directly via maximizing likelihood of next word given the ground truth word, we directly optimize the evaluation metrics such as BLEU \cite{papineni2002bleu} for code summarization. We apply a critic network to approximate the value of generated action at time step $t$. Different from the actor network, this critic network outputs a single value instead of a probability distribution on each decoding step. Before introducing critic network, we introduce the value function.

Given the policy $\pi$, sampled actions and reward function, the value function $V^{\pi}$ is defined as the prediction of total reward from the state $\mathbf{s}_t$ at step $t$ under policy $\pi$, which is formulated as follows:

\begin{equation}
V^{\pi}(\mathbf{s}_t)=\mathbb{E}_{\overset{\mathbf{s}_{t+1:T},}{y_{t:T}}}\left [\sum_{l=0}^{T-t}r_{t+l}|y_{t+1},\cdots,y_{T}, \mathbf{h} \right],
\end{equation}
where $T$ is the max step of decoding; $\mathbf{h}$ is the representation of code snippet. %which is the  calculated by the AST-based LSTM proposed in the actor network., at the cost of introducing bias. 
For code summarization, we can only obtain an evaluation score (BLEU) when the sequence generation process (or episode) is finished. The episode terminates when step exceeds the max-step $T$ or generating the end-of-sequence (EOS) token. Therefore, we define the reward as follows:

\begin{equation}
r_t=\left\{\begin{matrix}
0 & t < T \\ 
BLEU & t = T\ or\ EOS
\end{matrix}\right.
.
\end{equation}

Mathematically, the critic network tries to minimize the following loss function, where mean square error is used.

\begin{equation}
\mathcal{L}(\phi) = \frac{1}{2}\left \| V^{\pi}(\mathbf{s}_t) - V_\phi^\pi(\mathbf{s}_t) \right \|^2,
\end{equation}
where $V^{\pi}(\mathbf{s}_t) $ is the target value, $V_\phi^\pi(\mathbf{s}_t)$ is the value predicted by critic network and $\phi$ is the parameter of critic network.
\subsection{Model Training}
We use the policy gradient method to optimize policy directly, which is widely used in reinforcement learning. For actor network, the goal of training is to minimize the negative expected reward, which can be defined as $\mathcal{L}(\theta) =- \mathbb{E}_{y_{1,\ldots,T}\sim \pi}(\sum_{l=t}^{T}r_t)$, where $\theta$ is the parameter of actor network. 
Denote all the parameters as $\Theta=\{\theta, \phi \}$, the total loss of our model can be represented as $\mathcal{L}(\Theta)=\mathcal{L}(\theta)+ \mathcal{L}(\phi)$.

For policy gradient, it is typically better to train an expression of the following form according to \cite{schulman2015high}:

\begin{equation} \label{eq_gd_actor}
\nabla_{\theta}\mathcal{L}(\Theta)=\mathbb{E}[\sum_{t=0}^{T-1}A^{\pi}(\mathbf{s}_t,y_{t+1})\nabla_{\theta}\log \pi_{\theta}(y_{t+1}|\mathbf{s}_t)],
\end{equation}
where $A^{\pi}(\mathbf{s}_t,y_{t+1})$ is advantage function. The reason why we choose advantage function is that it achieves smaller variance when compared with some other ones such as TD residual and reward with baseline \cite{schulman2015high}. 

According to the definition of advantage function, we can formulate the advantage function as follows. One can refer to \cite{schulman2015high} for more details.

\begin{equation}\label{eq_reward}
A^{\pi}(\mathbf{s}_t,y_{t}) = Q^{\pi}(\mathbf{s}_t,y_{t}) - V^{\pi}(\mathbf{s}_t),
\end{equation}
where $Q^{\pi}(\mathbf{s}_t,y_{t})$ is the state-action value function which is defined as $Q^{\pi}(\mathbf{s}_t,y_{t}) = \mathbb{E}_{\overset{\mathbf{s}_{t+1:T},}{y_{t+1:T}}}\left [ \sum_{l=0}^{T-t}r_{t+l} \right ]$. 
From this formulation, we can find that the advantage function measures whether or not the action is better or worse than the policy's default behavior. Therefore, a step in the policy gradient direction can increase the probability of better-than-average actions and decrease the probability of worse-than-average actions.

On the other hand, the gradient of critic network is calculated as follows: 

\begin{equation}\label{eq_gd_critic}
\nabla_{\phi}\mathcal{L}(\Theta)=\sum_{t=0}^{T-1}[V^\pi(\mathbf{s}_t)-V_\phi^\pi(\mathbf{s}_t)]\nabla_{\phi}V_\phi^\pi(\mathbf{s}_t).
\end{equation}

We employ stochastic gradient descend with the diagonal variant of AdaGrad \cite{duchi2011adaptive} to optimize the parameters of our framework. 
Algorithm 1 summarizes our proposed model described above.

\begin{algorithm}[!t]
	\caption{Actor-Critic training for code summarization.}\label{alg:ac-training}
	\begin{algorithmic}[1]
		\STATE Initialize actor $\pi_{y_{t+1|s_t}}$ and critic $V^{\pi}(s_t)$ with random weights $\theta$ and $\phi$;
		\STATE Pre-train the actor to predict ground truth $y_t$ given $\{ y_1, \cdots, y_{t-1} \}$ by minimizing Eq. \ref{eq_actor_objective};
		\STATE Pre-train the critic to estimate $V(s_t)$ with fixed actor;
		
		\FOR{$t=1\to T$}
		\STATE Receive a random example, and generate sequence of actions $\{ y_1, \cdots , y_T\}$ according to current policy $\pi_{\theta}$;
		\STATE Calculate advantage estimate $A^{\pi}$ according to Eq. \ref{eq_reward};
		\STATE Update critic weights $\phi$ using the gradient in Eq. \ref{eq_gd_critic};
		\STATE Update actor weights $\theta$ using the gradient in Eq. \ref{eq_gd_actor}.
		\ENDFOR
	\end{algorithmic}
\end{algorithm}
	
	\section{Experiments and Analysis}\label{sec_experiments}
	To evaluate our proposed approach, in this section, we conduct experiments to answer the following questions:
	\begin{itemize}
		\item \textbf{RQ1.} Does our proposed approach improve the performance of code summarization when compared with some state-of-the-art approaches?
		\item \textbf{RQ2.} What's the effectiveness of each component for our proposed model? For example, what about the performance of hybrid code representation and reinforcement learning respectively?
		\item \textbf{RQ3.} What's the performance of our proposed model on the datasets with different code or comment length?
	\end{itemize}
	
	We ask RQ1 to evaluate our deep reinforcement learning-based model compared to some state-of-the-art baselines. %which will be describe in the following subsection. 
	We ask RQ2 in order to evaluate each component of our model. We ask RQ3 to evaluate our model when varying the length of code or comment. In the following subsections, we first describe the dataset, some evaluation metrics and the training details. Then, we introduce some baselines for RQ1. Finally, we report our results and analysis for the research questions.
	
	\subsection{Dataset Preparation}
	We evaluate the performance of our proposed method using the dataset in \cite{barone2017parallel}, which is obtained from a popular open source projects hosting platform, GitHub\footnote{https://github.com/}. The dataset contains 108,726 code-comment pairs. The vocabulary size of code and comment is 50,400 and 31,350, respectively. For cross-validation, 
	We shuffle the dataset and use the first 60\% for training, 20\% for validation and the remaining for testing. To construct the tree-structure of code, we parse Python code into abstract syntax trees via ast\footnote{https://docs.python.org/2/library/ast.html} lib. To convert code into sequential text, we tokenize the code by \{. , " ' : ; ) ( !  (space)\}, which has been used in \cite{nguyen2017automatic}. We tokenize the comment by \{(space)\}. 
	
\begin{figure}[!t]
		\centering
		\begin{subfigure}[b]{1.65in}
			\includegraphics[width=\textwidth]{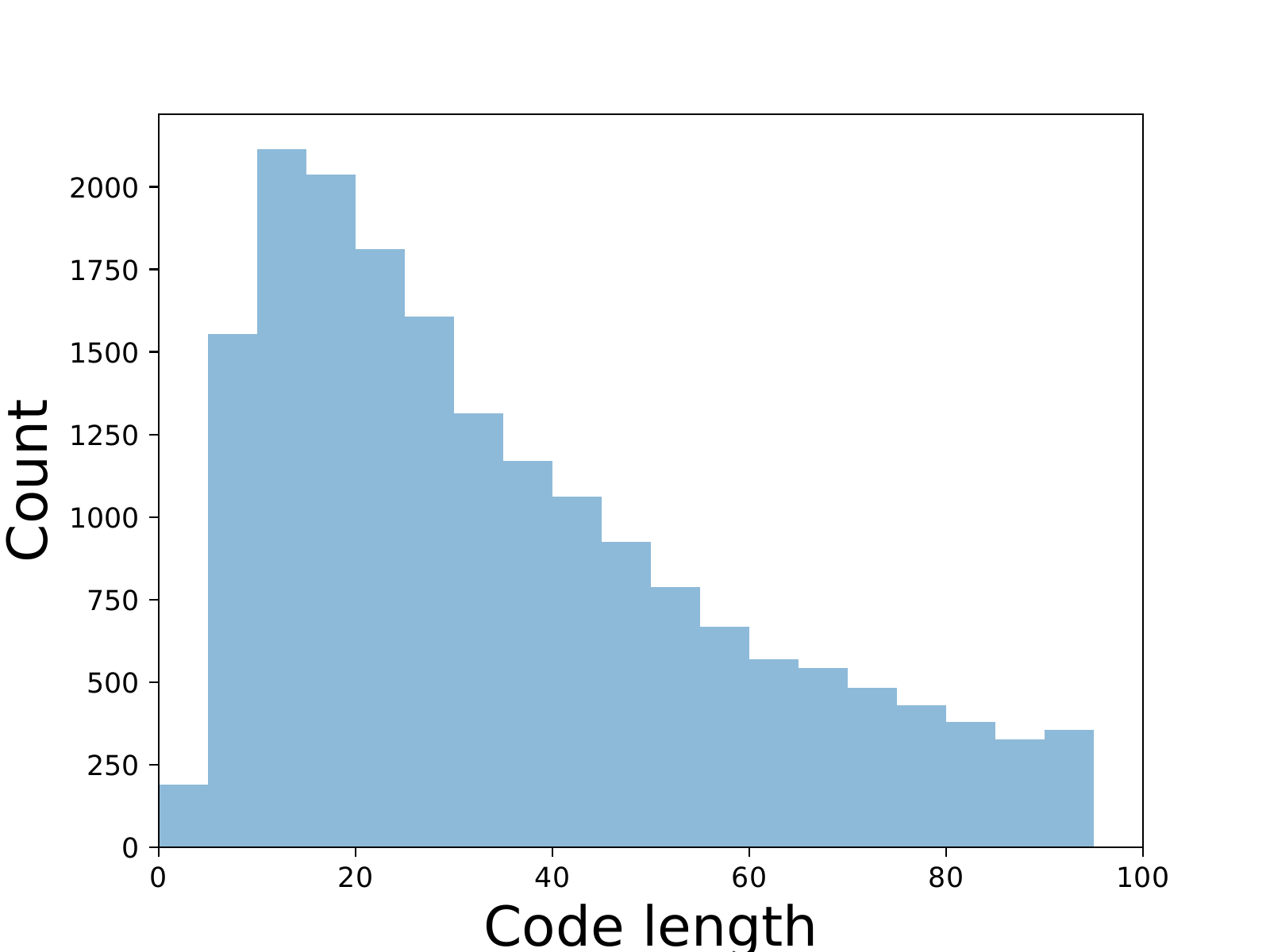}
			\caption{Code length distribution.}
			\label{fig_code_length}
		\end{subfigure}
		\begin{subfigure}[b]{1.65in}
			\includegraphics[width=\textwidth]{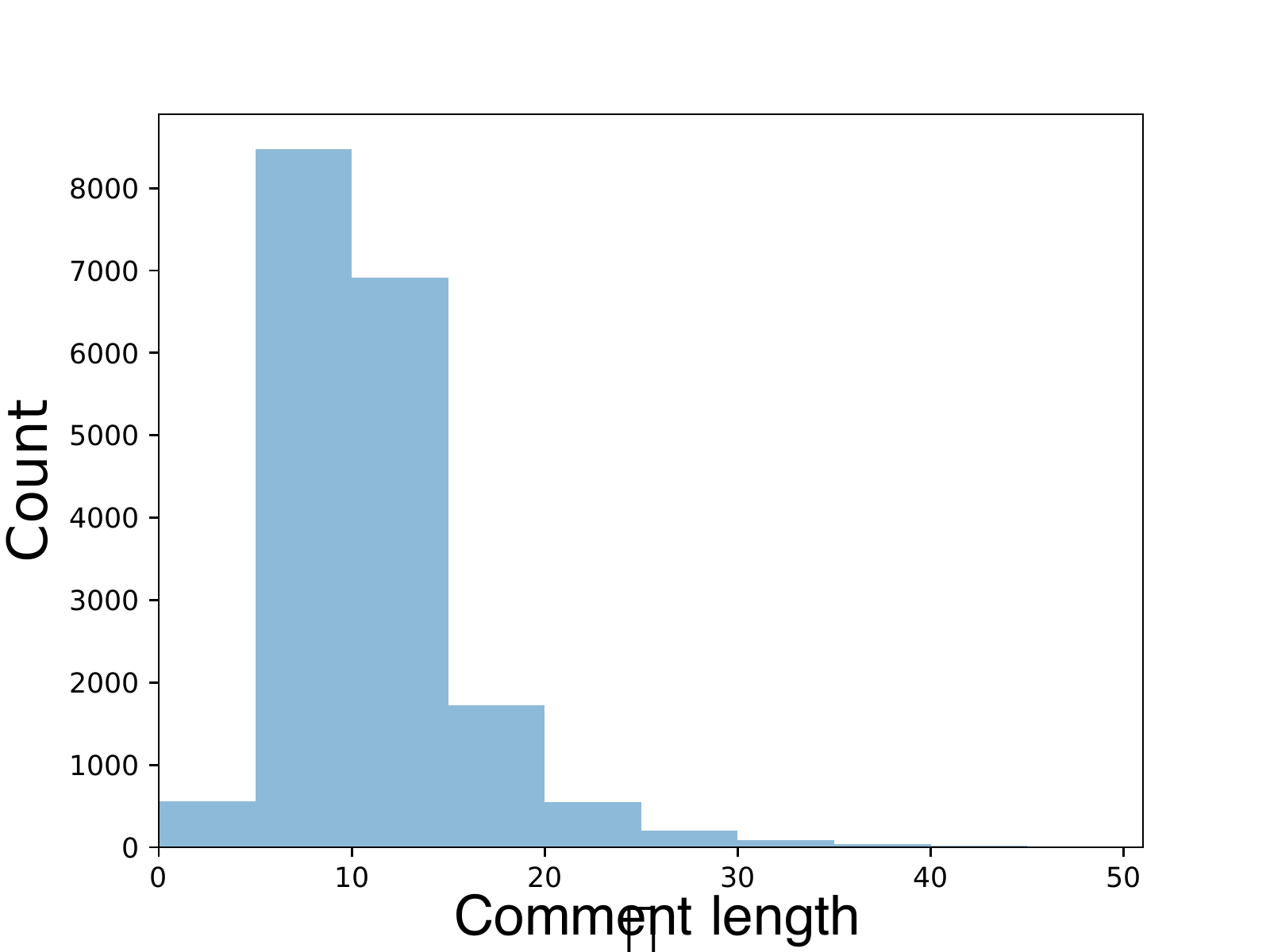}
			\caption{Comment length distribution.}
			\label{fig_comment_length}
		\end{subfigure}
		\caption{Length distribution of testing data.}
		\label{fig_length_distribution}
	\end{figure}

	Figure \ref{fig_length_distribution} shows the length distribution of code and comment on testing data. From Figure \ref{fig_code_length}, we can find that the lengths of most code snippets are located between 20 to 60. This verifies the quote in \cite{martin2009clean} ``Functions should hardly ever be 20 lines long". In Python language, the limited length should be shorter. From Figure \ref{fig_comment_length}, we can notice that the length of nearly all the comments are between 5 to 15. This reveals the comment sequence that we need to generate will not be too long.
	
	\subsection{Evaluation Metrics}
	We evaluate the performance of our proposed model based on four widely-used evaluation criteria in the area of neural machine translation and image captioning, i.e., BLEU \cite{papineni2002bleu}, METEOR \cite{banerjee2005meteor},  ROUGE-L \cite{lin2004rouge} and CIDER \cite{vedantam2015cider}. BLEU measures the average n-gram precision on a set of reference sentences, with a penalty for short sentences. METEOR is recall-oriented and measures how well our model captures content from the references in our output. ROUGE-L takes into account sentence level structure similarity naturally and identifies longest co-occurring in sequence n-grams automatically. CIDER is a consensus based evaluation protocol for image captioning. %To make the paper be compact, we put the formulation of each metric in Table \ref{table_metric} (see Appendix B).
	\subsection{Training Details}
\begin{figure}[!t]
		\centering
		\includegraphics[width=0.46\textwidth]{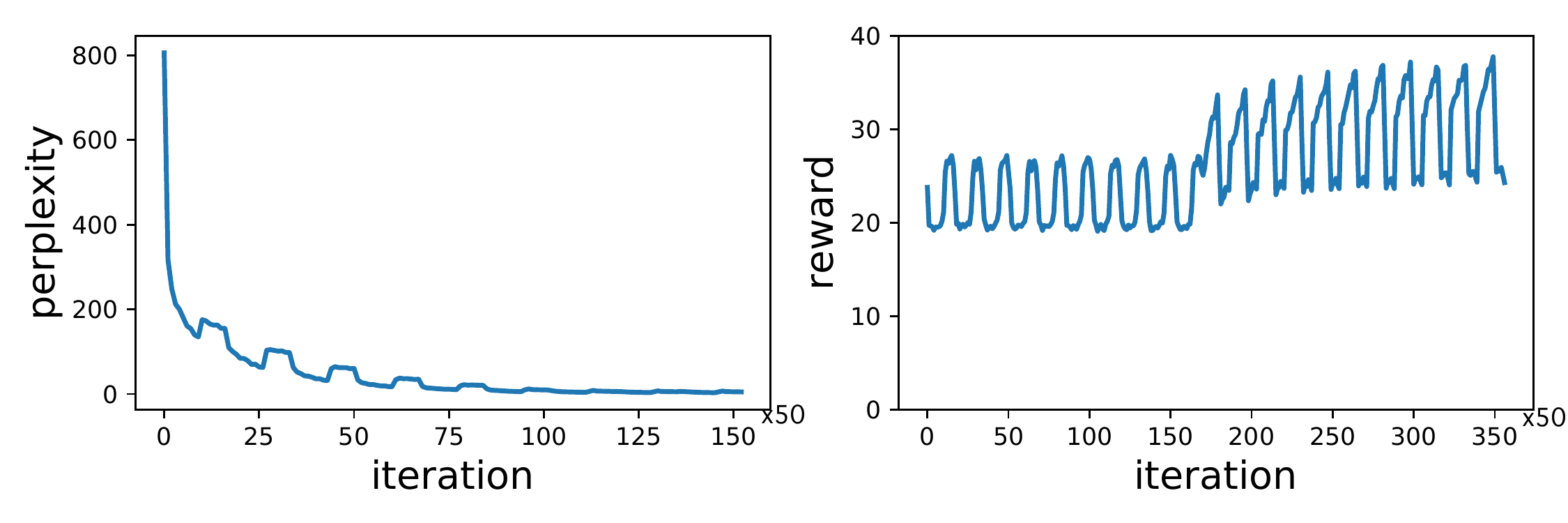}
		\caption{Iteration of training perplexity and reward.}
		\label{fig_training_curve}
	\end{figure}

	The hidden size of the encoder and decoder LSTM networks are both set to be 512, and the word embedding size is set to be 512. The mini-batch size is set to be 64, while the learning rate is set to be $0.001$. We pretrain both actor network and critic network with 10 epochs each, and train the actor-critic network simultaneously 10 epoches. We record the perplexity\footnote{Perplexity is a function of cross entropy loss, which has been widely used in evaluation of many natural language processing tasks.}/reward every 50 iterations. Figure \ref{fig_training_curve} shows the perplexity and reward curves of our method. All the experiments in this paper are implemented with Python 2.7, and run on a computer with an 2.2 GHz Intel Core i7 CPU, 64 GB 1600 MHz DDR3 RAM, and a Titan X GPU with 12 GB memory, running Ubuntu 16.04. 
	
	\subsection{RQ1: Compared to Baselines}
	\begin{table*}[!t]
		\centering
		\caption{Comparison of the overall performance between our model and previous methods. (Best scores are in boldface.)}
		\label{table_overall_performance}
		\begin{tabular}{|l|l|l|l|l|l|l|l|}
			\hline
			& \textbf{BLEU-1} &\textbf{BLEU-2} & \textbf{BLEU-3} & \textbf{BLEU-4} & \textbf{METEOR} & \textbf{ROUGE-L} & \textbf{CIDER} \\
			\hline
			Seq2Seq      &0.1660	&0.0251	&0.0100	&0.0056	&0.0535	&0.2838	&0.1262        \\
			Seq2Seq+Attn & 0.1897	&0.0419	&0.0200	&0.0133	&0.0649	&0.3083	&0.2594          \\
			Tree2Seq      & 0.1649	&0.0236	&0.0096	&0.0053	&0.0501	&0.2794	&0.1168         \\
			Tree2Seq+Attn & 0.1887	&0.0417	&0.0197	&0.0129	&0.0644	&0.3068	&0.2331          \\
			Hybrid2Seq+Attn+DRL (Our)   & \textbf{0.2527}	&\textbf{0.1033 }&\textbf{0.0640}	&\textbf{0.0441}	&\textbf{0.0929}	&\textbf{0.3913}	&\textbf{0.7501}       \\
			\hline    
		\end{tabular}
	\end{table*}
	
	\begin{table*}[!t]
		\centering
		\caption{Effectiveness of each component for our proposed model. (Best scores are in boldface.)}
		\label{table_component_performance}
		\begin{tabular}{|l|l|l|l|l|l|l|l|}
			\hline
			& \textbf{BLEU-1} &\textbf{ BLEU-2} & \textbf{BLEU-3} & \textbf{BLEU-4} & \textbf{METEOR} & \textbf{ROUGE-L} & \textbf{CIDER} \\
			\hline
			Seq2Seq+Attn+DRL      &0.2421	&0.0919	&0.0513	&0.0325	&0.0882	&\textbf{0.3935}	&0.6390     \\
			Tree2Seq+Attn+DRL      & 0.2309	&0.0854	&0.0499	&0.0338	&0.0843	&0.3767	&0.6060     \\
			Hybrid2Seq      & 0.1837	&0.0379	&0.0183	&0.0122	&0.0604	&0.3020	&0.2223     \\
			Hybrid2Seq+Attn & 0.1965	&0.0516	&0.0280	&0.0189	&0.0693	&0.3154	&0.3475       \\
			Hybrid2Seq+Attn+DRL (Our)    & \textbf{0.2527}	&\textbf{0.1033 }&\textbf{0.0640}	&\textbf{0.0441}	&\textbf{0.0929}	&0.3913	&\textbf{0.7501}       \\
			\hline    
		\end{tabular}
	\end{table*}
	%	To verify the effectiveness our proposed model, w
	We compare our model with the following baselines:
	\begin{compactitem}
		\item Seq2Seq \cite{sutskever2014sequence} is a classical encoder-decoder framework in neural machine translation, which encodes the source sentences into a hidden space, and decodes it into target sentences. In our comparison, the encoder and decoder are both based on LSTM.
		\item Seq2Seq+Attn \cite{bahdanau2014neural} is a derived version of Seq2Seq model with an attentional layer for word alignment.
		\item Tree2Seq \cite{weisupervised} follows the same architecture with Seq2Seq and applies AST-based LSTM as encoder for the task of code clone detection.
		\item Tree2Seq+Attn \cite{eriguchi2016tree}  is a derived version of Tree2Seq model with an attentional layer, which has been applied in neural machine translation
		\item Hybrid2Seq(+Attn+DRL) represents three versions of our proposed model with/without Attn/DRL component.
	\end{compactitem}
	
	Table \ref{table_overall_performance} shows the experimental results of comparison between our proposed model and some previous ones. From this table, we can find that our proposed model outperforms other baselines in almost all of evaluation metrics. When comparing Seq2Seq/Tree2Seq with its correspond attention-based version, we can see that attention is really effective in aligning the code tokens with comment tokens. We can also find that the performance of simply encoding the tree structure of code is worse than that of simply encoding the code as sequence. This can be illustrated by that the words of comments are always drawn from the tokens of code. Thus, our model which considers both the structure and sequential information of code achieves the best performance in this comparison.
	
	\subsection{RQ2: Component Analysis}
	%\subsection{RQ2: Performance of Hybrid Code Representation and Reinforcement Learning}
	Table \ref{table_component_performance} shows the effectiveness of some main components in our proposed model. From this table, comparing the results of Seq2Seq+ Attn/Tree2Seq+Attn with and without (Table \ref{table_overall_performance}) deep reinforcement learning (DRL), we can see that the proposed DRL component can really boost the performance of comment generation for source code. We can also find the proposed approach of integrating the LSTM for content and AST-based LSTM for structure is effective on representing the code as compared with the corresponding non-hybrid ones in Table \ref{table_overall_performance}. Furthermore, it also verifies that our proposed hybrid attention mechanism works well in our model.
	
	\subsection{RQ3: Parameter Analysis}
	\begin{figure*}[!t]
		\centering
		\begin{subfigure}[b]{1.7in}
			\includegraphics[width=\textwidth]{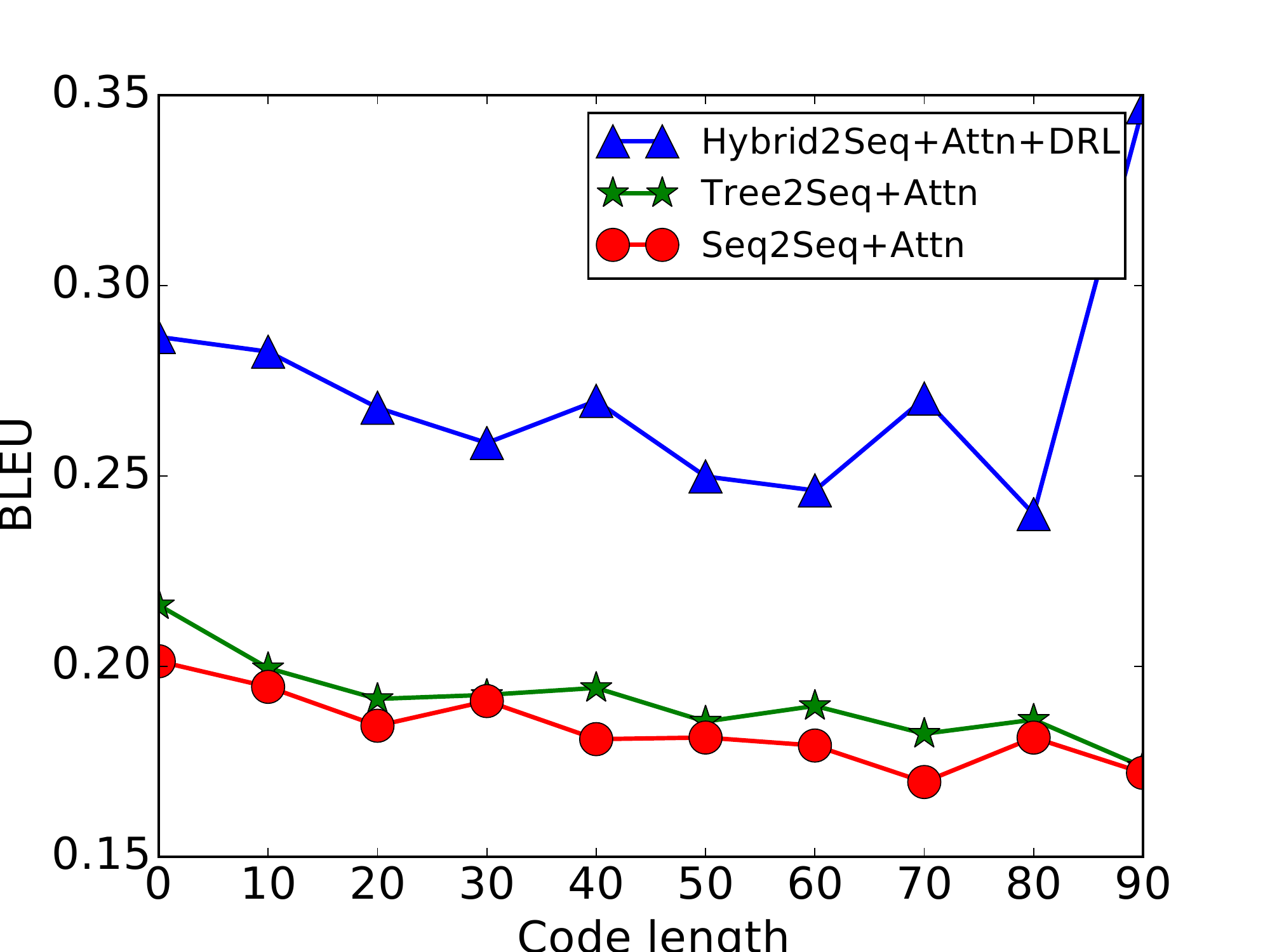}
			\caption{BLEU}
			\label{fig_varcode_bleu}
		\end{subfigure}
		\begin{subfigure}[b]{1.7in}
			\includegraphics[width=\textwidth]{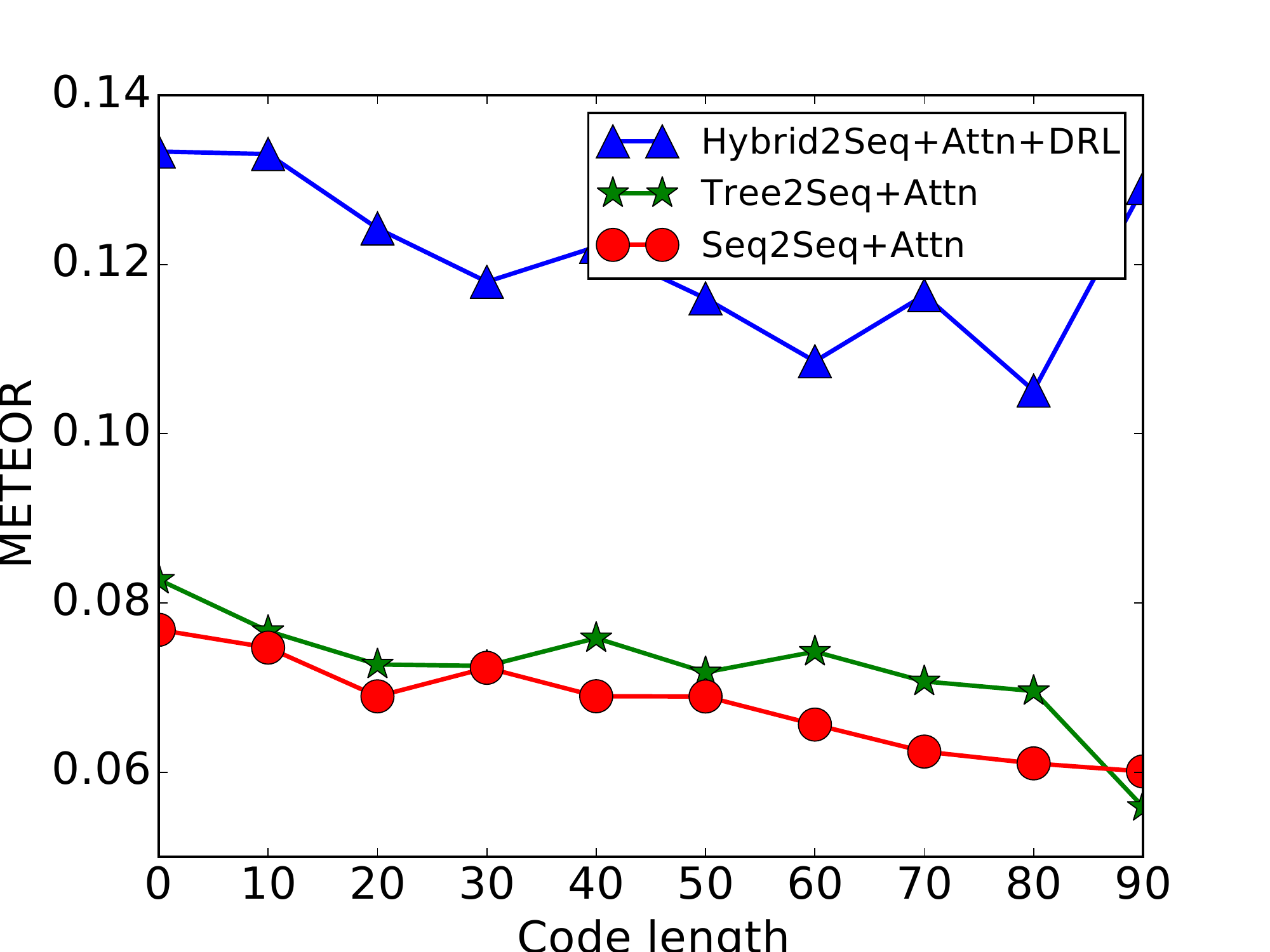}
			\caption{METEOR}
			\label{fig_varcode_meteor}
		\end{subfigure}
		\begin{subfigure}[b]{1.7in}
			\includegraphics[width=\textwidth]{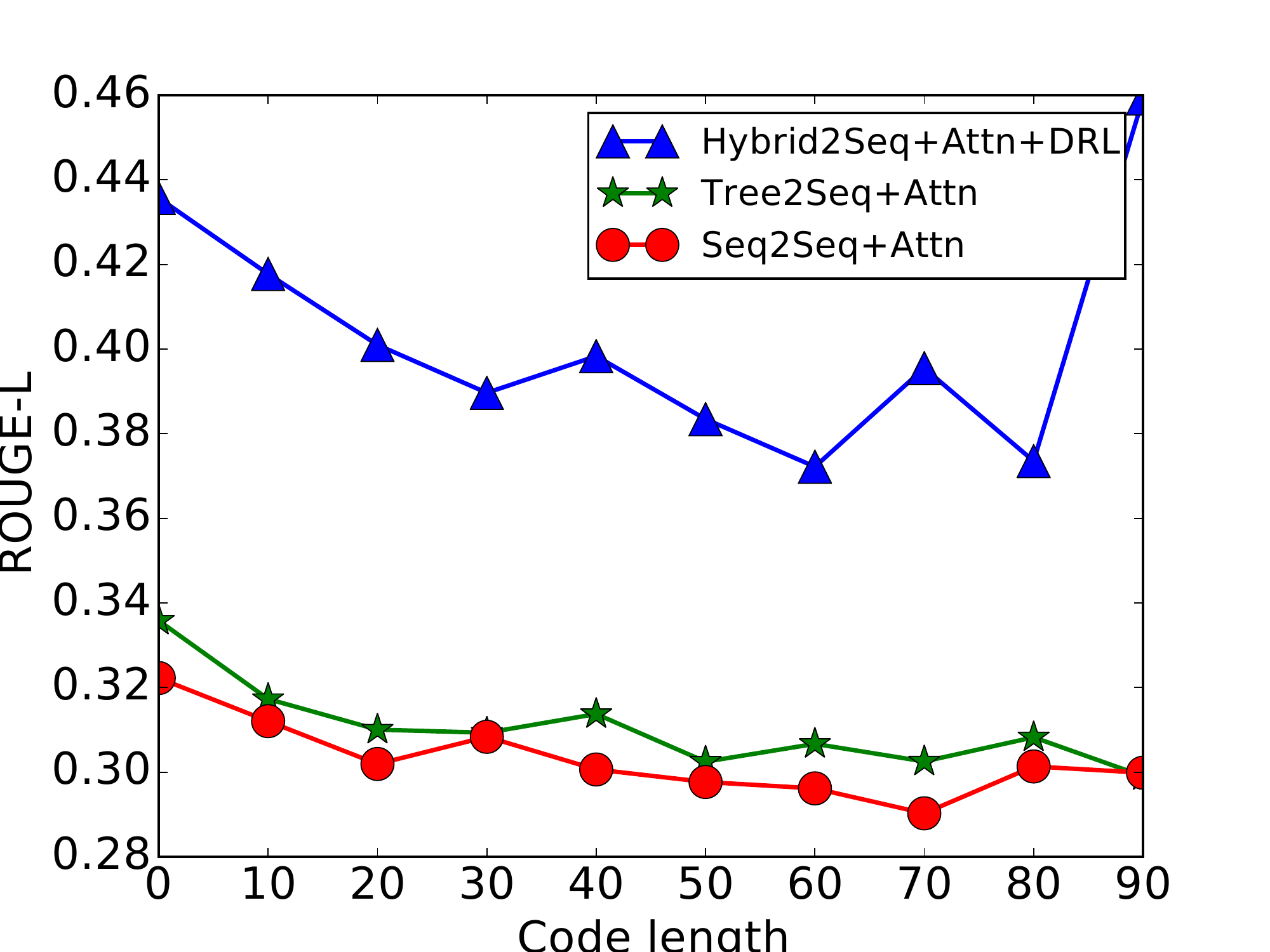}
			\caption{ROUGE-L}
			\label{fig_varcode_rouge}
		\end{subfigure}
		\begin{subfigure}[b]{1.7in}
			\includegraphics[width=\textwidth]{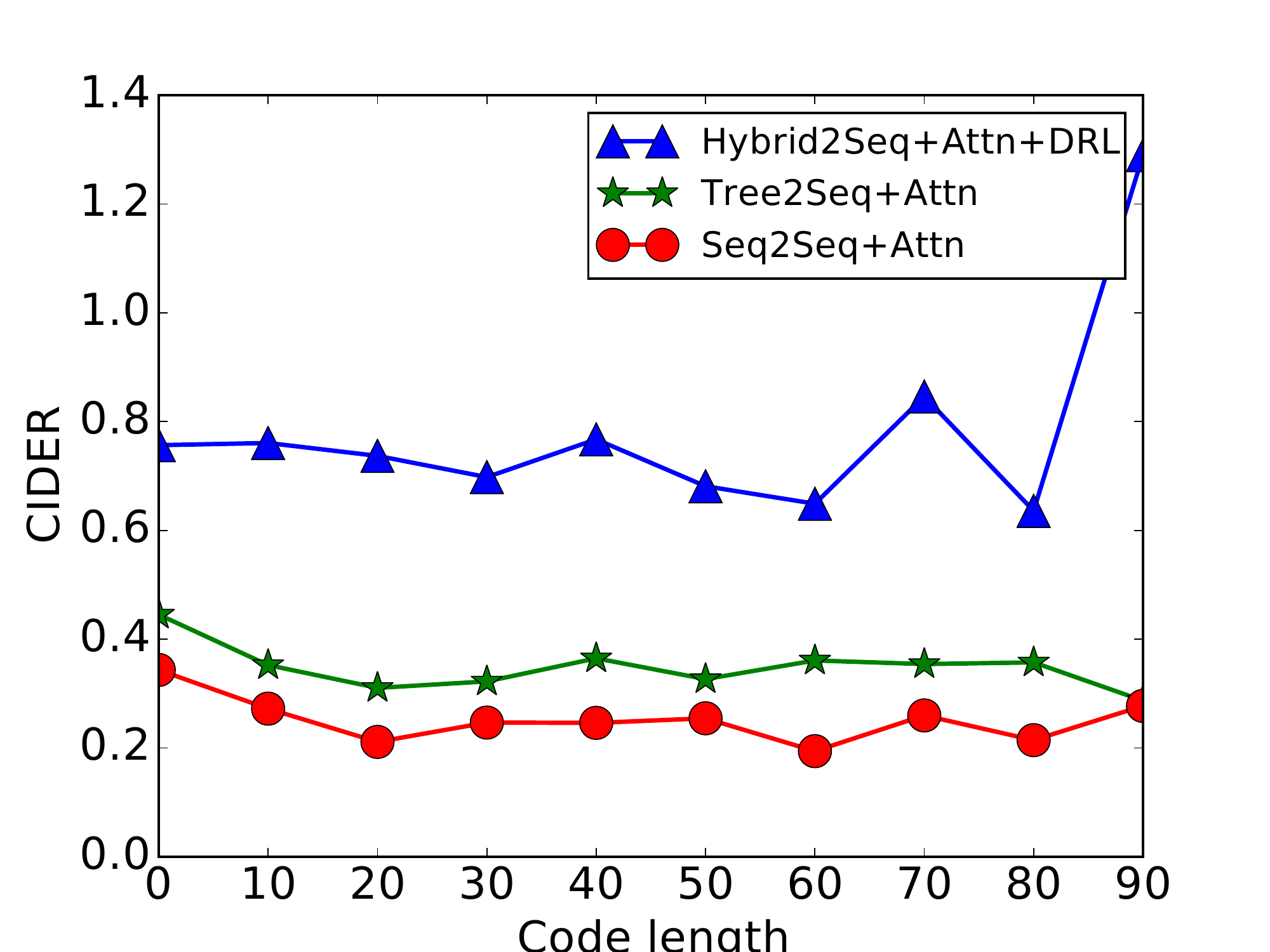}
			\caption{CIDER}
			\label{fig_varcode_cider}
		\end{subfigure}
		\caption{Experimental results of our proposed method and some baselines on different metrics w.r.t. varying code length.}
		\label{fig_varcode}
%		\vspace{-0.8em}
	\end{figure*}
	
	\begin{figure*}[!t]
		\centering
		\begin{subfigure}[b]{1.7in}
			\includegraphics[width=\textwidth]{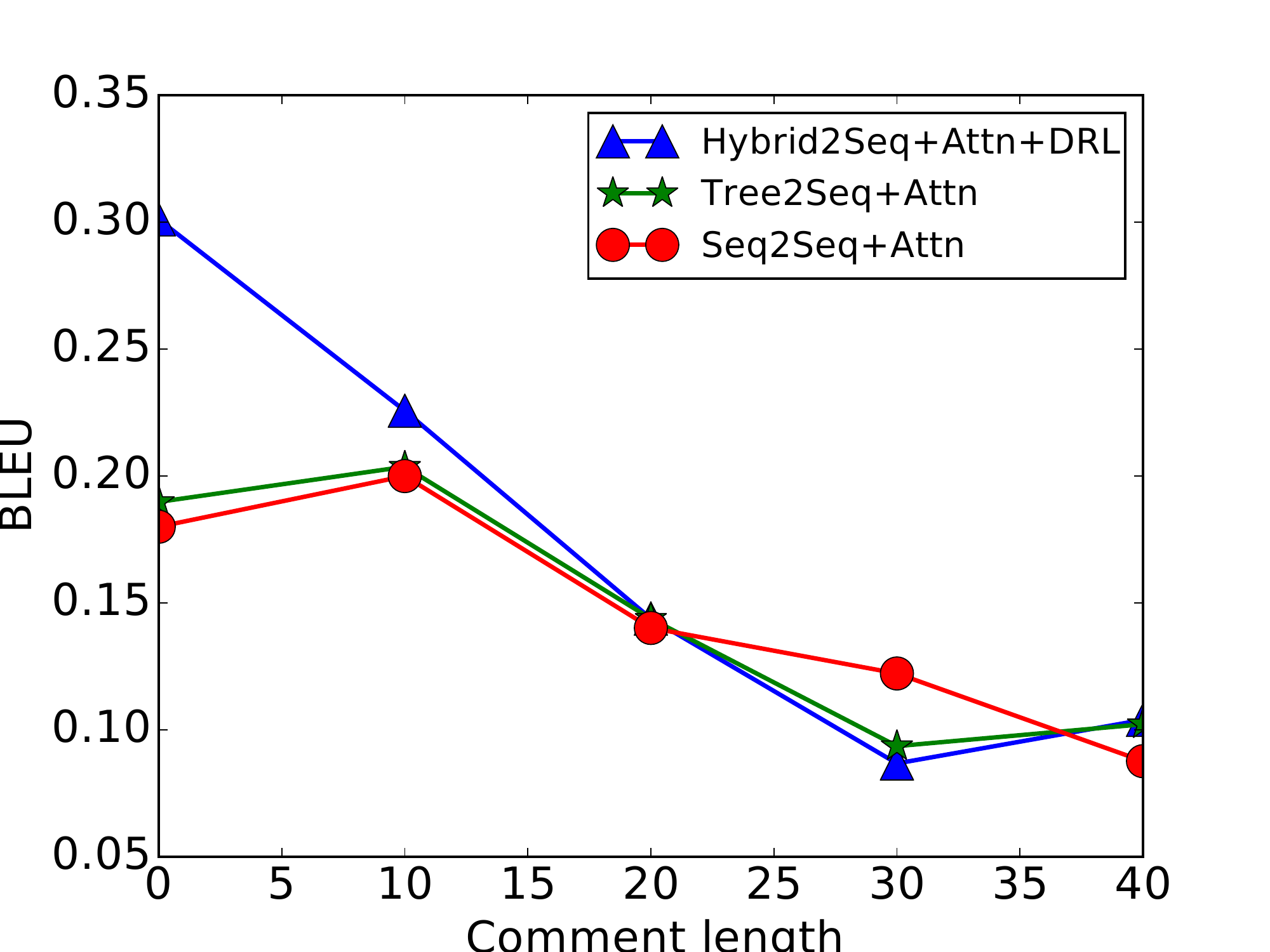}
			\caption{BLEU}
			\label{fig_varcomment_bleu}
		\end{subfigure}
		\begin{subfigure}[b]{1.7in}
			\includegraphics[width=\textwidth]{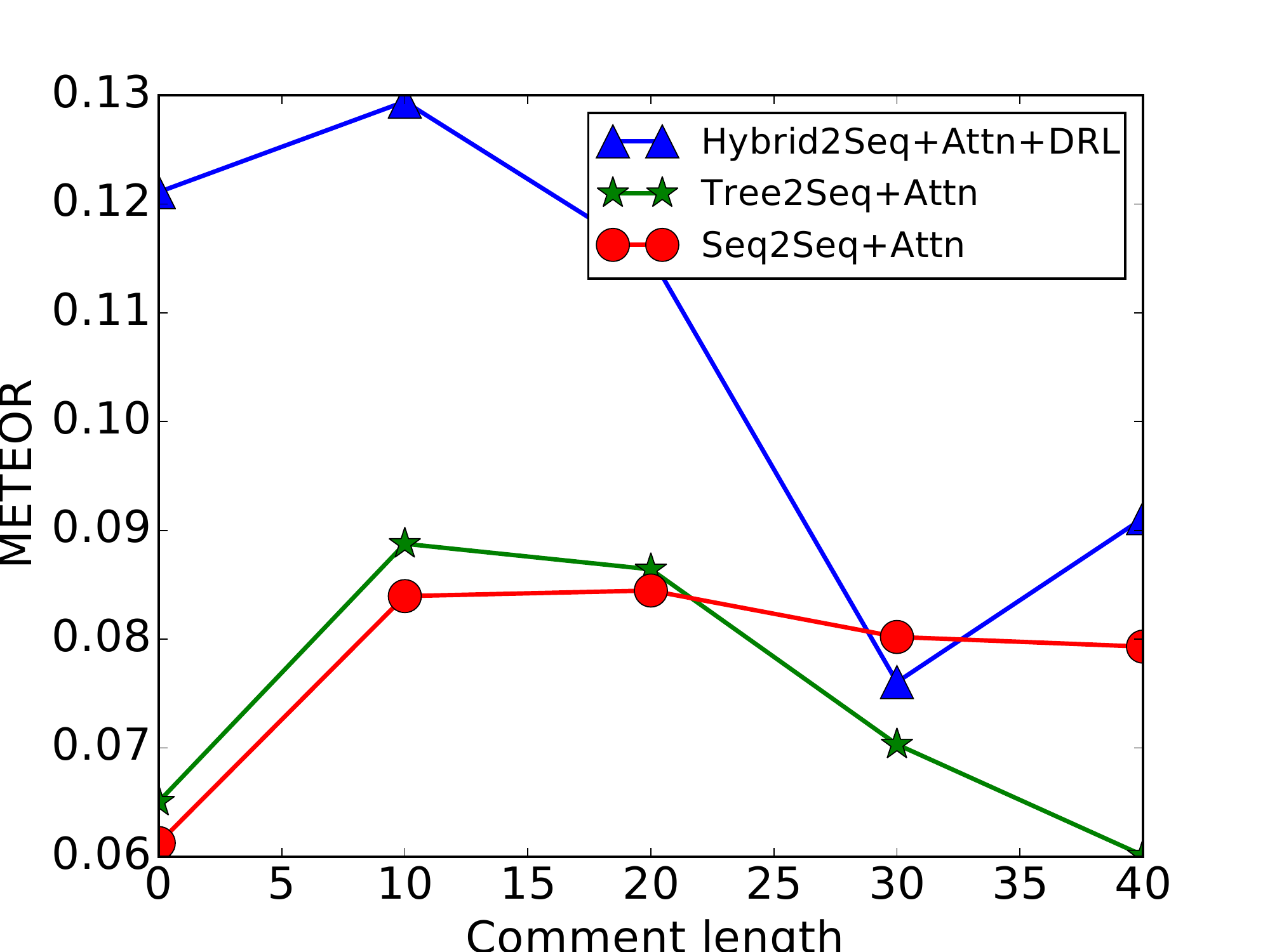}
			\caption{METEOR}
			\label{fig_varcomment_meteor}
		\end{subfigure}
		\begin{subfigure}[b]{1.7in}
			\includegraphics[width=\textwidth]{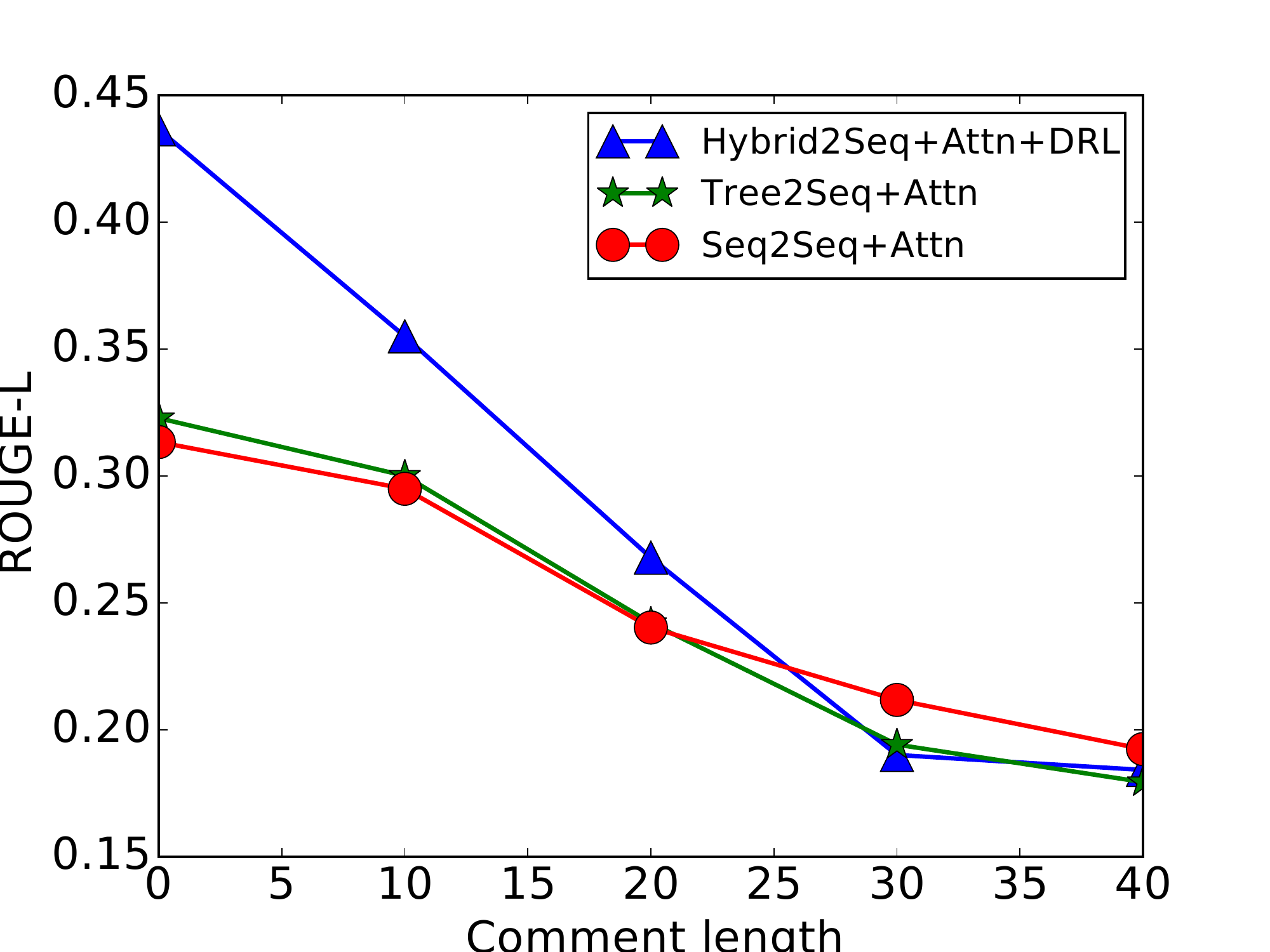}
			\caption{ROUGE-L}
			\label{fig_varcomment_rouge}
		\end{subfigure}
		\begin{subfigure}[b]{1.7in}
			\includegraphics[width=\textwidth]{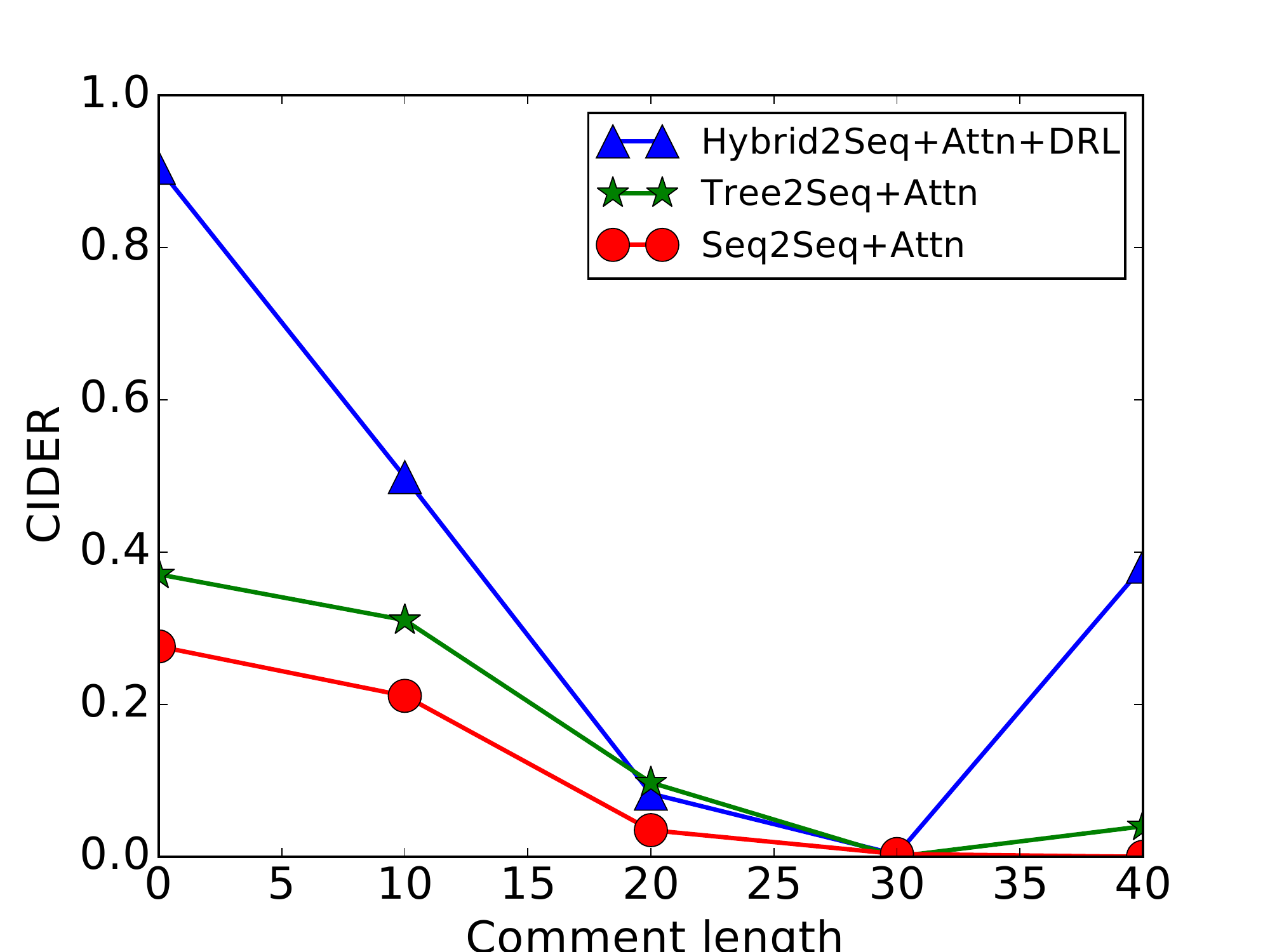}
			\caption{CIDER}
			\label{fig_varcomment_cider}
		\end{subfigure}
		
		\caption{Experimental results of our proposed method and some baselines on different metrics w.r.t. varying comment length.}
		\label{fig_varcomment}
%		\vspace{-0.8em}
	\end{figure*}
	We vary the length of code and comment since the code length may have an effect on the representation of code and the comment length may have an effect on the performance of text generation. 
	%	We vary the length of comment since the performence of text generation varies a lot on different length of text.
	Figure \ref{fig_varcode} and Figure \ref{fig_varcomment} show the performance of our proposed method when compared with two baselines on datasets of varying code lengths and comment lengths, respectively. 
	
	From Figure \ref{fig_varcode}, we can see that our model performs best when compared with other baselines on four metrics with respect to different code lengths. Additionally, we can see that the our proposed model has a stable performance even though the code length increases dramatically. We attribute this effect to the hybrid representation we adopt in our model. 
	For Figure \ref{fig_varcomment}, recall the comment length distribution in Figure \ref{fig_comment_length}. Since nearly all the comment lengths of testing data are under 20, we ignore the performance analysis over the samples whose comment length are larger than $20$. From this figure, we can see the performances of our model and baselines vary dramatically on four metrics with respect to different comment lengths. 

	\subsection{Qualitative Analysis and Visualization}
	We show two examples in Table \ref{table_case_study}. It's clear that the generated comments by our model are closest to the ground truth. 
	Although those models without DRL can generate some tokens which are also in the ground truth, they can't predict those tokens which are not frequently appeared in the training data. On the contrary, our deep reinforcement learning based model can generate some tokens which are closer to the ground truth, like \texttt{git}, \texttt{symbolic}.
	This can be illustrated by the fact that our model has a more comprehensive exploration on the word space and optimizes the BLEU score directly.
	
	In Table \ref{table_case_study}, we also visualize two attentions in our proposed model for the target sentences. For example, for Case 1 with target sentence \textit{check if git is installed .}, we can notice that the str-attn (left of figure) focuses more on tokens like \texttt{OSError}, \texttt{False}, \texttt{git}, \texttt{version}, which represent the structure of code. On the other hand, the attention of txt-attn (right of figure) is comparatively dispersed, and have a focus on some tokens like \texttt{def}, which is of little significance for code summarization. This verifies our assumption that LSTM can capture the sequential content of code, and AST-based LSTM can capture the structure information of code. Thus, it's reasonable to fuse them together for a better representation.
	%the left code-attention focuses more on the structure of code, while the right text-attention focused on the content of code. This verifies our assumption that LSTM can contribute to the textual information for comment generation, and Tree-LSTM can capture the structure information of code snippets. 
	
	\section{Threats to Validity and Limitations}\label{sec_threats}
	One threat to validity is that our approach is experimented only on Python code collected from GitHub, so they may not be representative of all the comments. However, Python is a popular programming language used in a large number of projects. In the future, we will extend our approach to other programming languages. 
	Another threat to validity is on the metrics we choose for evaluation. It has always been a tough challenge to evaluate the similarity between two sentences for the tasks such as neural machine translation \cite{sutskever2014sequence}, image captioning \cite{kilickaya2016re}. In this paper, we only adopt four popular automatic metrics, it is necessary for us to evaluate the performance of generated text from more perspectives, such as human evaluation.
	Furthermore, in the deep reinforcement learning perspective, we only set the BLEU score of generated sentence as the reward. It's well known that for a reinforcement learning method, one of the biggest challenge is how to design a reward function to measure the value of action correctly, and it is still an open problem. In our future work, we plan to devise a reward function that can reflect the value of each action more correctly.
	\begin{table*}[!t]
		\centering
		\caption{Examples of code summarization generated by each model and attention visualization of our model. }
		\label{table_case_study}
		\begin{tabular}{|l|l|l|}
			\hline
			\textbf{}&\textbf{Case 1} &\textbf{Case 2}\\
			\hline
			Code snippet &
			\begin{tabular}
				[c]{@{}l@{}}
				\texttt{{\textcolor{code-blue}{def} \_has\_git():}} \\  \ \ 
				\texttt{\textcolor{code-blue}{try}: subprocess.check\_call(} \\  \ \  \ \ 
				%			\texttt{subprocess.check\_call(} \\   \ \  \ \  \ \ 
				\texttt{[git, --version], }\\   \ \  \ \  \ \ 
				\texttt{\textcolor{code-purple}{stdout}=subprocess.DEVNULL,} \\   \ \  \ \  \ \ 
				\texttt{\textcolor{code-purple}{stderr}=subprocess.DEVNULL)} \\  \ \ 
				\texttt{\textcolor{code-blue}{except}(OSError, subprocess}\\  \ \ \ \
				\texttt{.CalledProcessError):} \\  \ \ \ \
				\texttt{\textcolor{code-blue}{return False}} \\  \ \  
				\texttt{\textcolor{code-blue}{else: return True} }\\
				%			\texttt{\textcolor{code-blue}{return True} }
			\end{tabular} 
			&
			\begin{tabular}
				[c]{@{}l@{}}
				\texttt{\textcolor{code-blue}{def} tensor3(\textcolor{code-red}{name}=None,\textcolor{code-red}{dtype}=None): }\\  \ \ 
				\texttt{\textcolor{code-blue}{if} (\textcolor{code-red}{dtype} \textcolor{code-blue}{is} \textcolor{code-blue}{None}): }\\  \ \ \ \ 
				\texttt{dtype=config.floatX }\\  \ \ 
				\texttt{type=CudaNdarrayType(} \\  \ \   \ \   \ \   \ \   \ \   \ \ 
				\texttt{\textcolor{code-purple}{dtype}=\textcolor{code-red}{dtype}, }\\   \ \    \ \   \ \   \ \   \ \   \ \ 
				\texttt{\textcolor{code-purple}{broadcastable}= }\\  \ \   \ \   \ \   \ \   \ \   \
				\texttt{(\textcolor{code-blue}{False}, \textcolor{code-blue}{False}, \textcolor{code-blue}{False})) } \\  \ \ 
				\texttt{\textcolor{code-blue}{return} type(\textcolor{code-red}{name}) }
			\end{tabular} \\
			\hline
			Ground truth & check if git is installed . & return a symbolic 3-d variable .\\
			\hline
			Seq2Seq & \begin{tabular}
				[c]{@{}l@{}}helper function to create a new figure\\ manager instance .\end{tabular} & yaml \\
			\hline
			Seq2Seq+Attn & \begin{tabular}
				[c]{@{}l@{}}return true if the user has access to the\\ specified resource . \end{tabular} & \begin{tabular}
				[c]{@{}l@{}}a decorator that returns a new class that will return \\a new class name .\end{tabular}\\
			\hline
			Tree2Seq+Attn & \begin{tabular}
				[c]{@{}l@{}}test that validate\_folders throws a \\ foldermissingerror . \end{tabular}  & helper function for \#4957 .\\
			\hline
			Hybrid2Seq+Attn & \begin{tabular}
				[c]{@{}l@{}}returns the number of git modules that are \\ not installed . \end{tabular}  & return the path to the currently running server .\\
			\hline
			Hybrid2Seq+Attn+DRL & \fbox{returns} \fbox{true} \underline{if} \underline{git} \underline{is} \underline{installed} . & \underline{return} \underline{a} \underline{symbolic} graph .\\
			\hline
			Attention visualization&\includegraphics[width=0.36\textwidth]{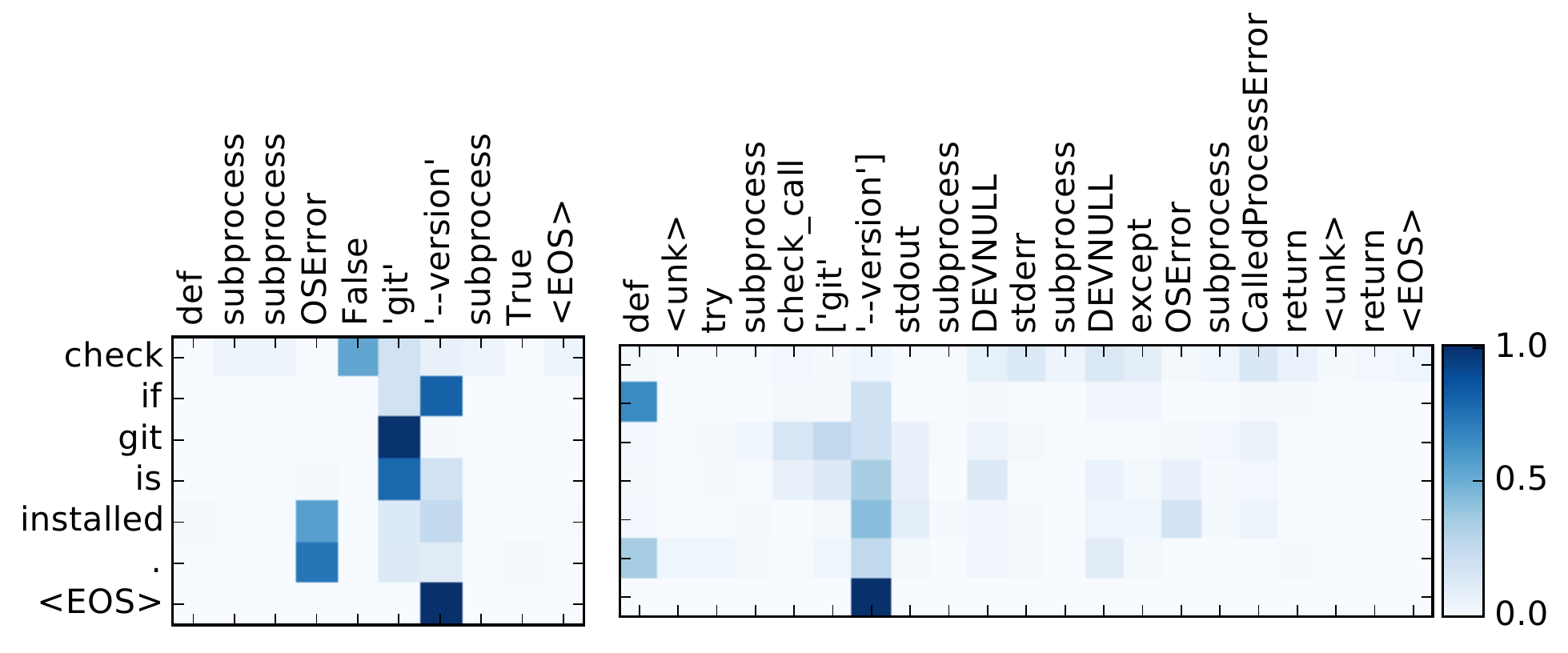}&\includegraphics[width=0.40\textwidth]{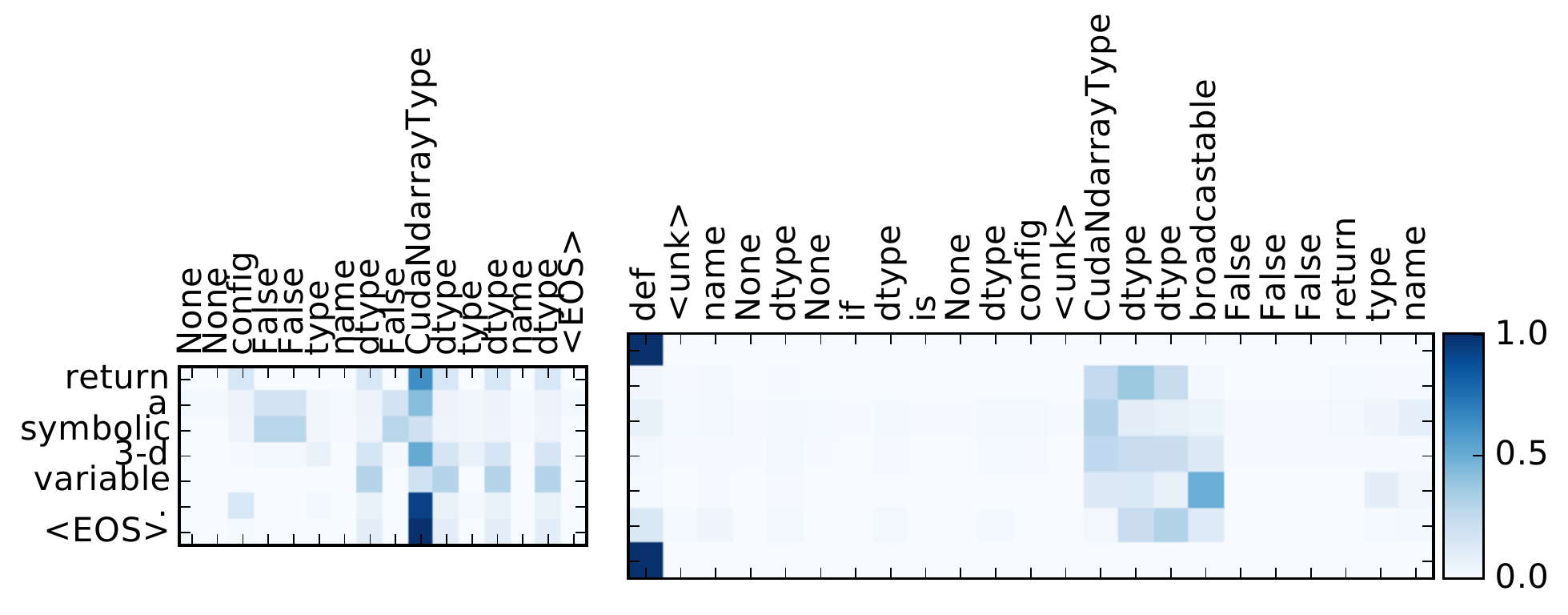}\\
			\hline
		\end{tabular}
	\end{table*}

	\section{Related Work}\label{sec_relatedwork}
%	In this section, we briefly review some related work from the aspects of deep learning on code analysis, source code summarization and deep reinforcement learning.
%on some related tasks such as neural machine translation and image captioning.
\subsection{Deep Code Representation}
With the successful development of deep learning, it has also become more and more prevalent for representing source code in the domain of software engineering research.
%Deep learning algorithms, as applied to software, automatically learn representations of software artifacts. 
Gu et al. \cite{gu2016deep} use a sequence-to-sequence deep neural network \cite{sutskever2014sequence}, originally introduced for statistical machine translation, to learn intermediate distributed vector representations of natural language queries which they use to predict relevant API sequences. Mou et al. \cite{mou2016convolutional} learn distributed vector representations using custom convolutional neural networks to represent features of snippets of code, then they assume that student solutions to various coursework problems have been intermixed and seek to recover the solution-to-problem mapping via classification. Li et al. \cite{li2015gated} learn distributed vector representations for the nodes of a memory heap and use the learned representations to synthesize candidate formal specifications for the code that produces the heap. 
%Li et al. \cite{li2015gated} exploit heap structure to define graph neural networks, a new machine learning model based on GRUs (a type of RNN) to directly learn from heap graphs. 
Piech et al. \cite{piech2015learning} and Parisotto et al. \cite{parisotto2016neuro} learn distributed representations of source code input/output pairs and use them to assess and review student assignments or to guide program synthesis from examples.
Neural code-generative models of code also use distributed representations to capture context, which is a common practice in natural language processing. For example, the work of Maddison and Tarlow \cite{maddison2014structured} and other neural language models (e.g. LSTMs in Dam et al. \cite{dam2016deep}) describe context distributed representations while sequentially generating code. Ling et al. \cite{ling2016latent} and Allamanis et al. \cite{allamanis2015bimodal} combine the code-context distributed representation with distributed representations of other modalities (e.g. natural language) to synthesize code. 
\subsection{Source Code Summarization}
%ref: Automatically Generating Commit Messages from Diffs using Neural Machine Translation
Code summarization is a novel task in the area of software engineering and has drawn great attention in recent years. The existing works for code summarization can be mainly categorized as rule based approaches \cite{sridhara2010towards}, statistical language model based approaches \cite{movshovitz2013natural} and deep learning based approaches \cite{allamanis2016convolutional,iyer2016summarizing,haije2016automatic}.
Sridhara et al. \cite{sridhara2010towards} construct a software word usage model first, and generate comment according to the tokenized function/variable names via rules. Movshovitz-Attias et al. \cite{movshovitz2013natural} predict comments from Java source files using topic models and n-grams. In \cite{allamanis2016convolutional}, the authors introduce an attentional neural network that employs convolution on the input tokens to detect local time-invariant and long-range topical attention features to summarize source code snippets into short, descriptive function name-like summaries. Iyer et al. \cite{iyer2016summarizing} propose to use LSTM networks with attention to produce sentences that describe C\# code snippets and SQL queries. In Haije's thesis \cite{haije2016automatic}, the code summarization problem is modeled as a machine translation task, and some translation models such as Seq2Seq \cite{sutskever2014sequence} and Seq2Seq with attention \cite{bahdanau2014neural} are employed. Unlike previous studies, we take the tree structure and sequential content of source code into consideration for a better representation of code. %and propose a deep reinforcement learning framework to generate code summarization.
\subsection{Deep Reinforcement Learning}
Reinforcement learning \cite{williams1992simple, konda2000actor, sutton2000policy}, concerned with how software agents ought to take actions in an environment so as to maximize the cumulative reward, is well suited for the task of decision-making. Recently, professional-level computer Go program has been designed by Silver et al. \cite{silver2016mastering} using deep neural networks and Monte Carlo Tree Search. Human-level gaming control \cite{mnih2015human} has been achieved through deep Q-learning. A visual navigation system \cite{zhu2017target} has been proposed recently based on actor-critic reinforcement learning model.
Text generation can also be formulated as a decision-making problem and there have been several reinforcement learning-based works on this specific tasks, including image captioning \cite{ren2017deep}, dialogue generation \cite{li2016deep} and sentence simplification \cite{zhang2017sentence}. Ren et al. \cite{ren2017deep} propose an actor-critic deep reinforcement learning model with an embedding reward for image captioning. Li et al. \cite{li2016deep} integrate a developer-defined reward with REINFORCE algorithm for dialogue generation.
In this paper, we follow an actor-critic reinforcement learning framework, while our focus is on encoding the structural and sequential information of code snippets simultaneously with an attention mechanism.

\section{Conclusion}\label{sec_conclusion}
In this paper, we first point out two issues (i.e., code representation and exposure bias) existing in traditional code summarization works. To handle these two issues, we first encode the structure and sequential content of code via AST-based LSTM and LSTM respectively. Then we add a hybrid attention layer to integrate them together. We then feed the code representation vector into a deep reinforcement learning framework, named actor-critic network. Comprehensive experiments on a real-world dataset show that our proposed model outperforms other competitive baselines and achieves state-of-the-art performance on several automatic metrics, namely BLEU, METEOR, ROUGE-L and CIDER. 
For future work, in the first place, we plan to design a copy mechanism to cope with rare words which are out of our vocabulary, and extend our experiments to other programming languages such as Java. Due to the low efficiency of LSTM, we also plan to apply some other networks such as convolutional neural network (CNN) for code representation.

\section*{Acknowledgments}
This work is partially supported by the Ministry of Education of China under grant of No.2017PT18, the Natural Science Foundation of China under grant of No. 61672453, 61773361, 61473273, 61602405, the WE-DOCTOR company under grant of No. 124000-11110 and the Zhejiang University Education Foundation under grant of No. K17-511120-017. This work is also supported by CCF-Tencent Open Research Fund, NSF through grants IIS-1526499, IIS-1763325, CNS-1626432, and NSFC 61672313.
	
	
	
	% http://mirror.ctan.org/biblio/ttex/contrib/doc/
	%\bibliographystyle{IEEEtran}
	%\begin{thebibliography}{1}
	%\end{thebibliography}

	\clearpage
	\bibliographystyle{IEEEtran}
	\bibliography{ref}

\begin{thebibliography}{10}
\providecommand{\url}[1]{#1}
\csname url@samestyle\endcsname
\providecommand{\newblock}{\relax}
\providecommand{\bibinfo}[2]{#2}
\providecommand{\BIBentrySTDinterwordspacing}{\spaceskip=0pt\relax}
\providecommand{\BIBentryALTinterwordstretchfactor}{4}
\providecommand{\BIBentryALTinterwordspacing}{\spaceskip=\fontdimen2\font plus
\BIBentryALTinterwordstretchfactor\fontdimen3\font minus
  \fontdimen4\font\relax}
\providecommand{\BIBforeignlanguage}[2]{{%
\expandafter\ifx\csname l@#1\endcsname\relax
\typeout{** WARNING: IEEEtran.bst: No hyphenation pattern has been}%
\typeout{** loaded for the language `#1'. Using the pattern for}%
\typeout{** the default language instead.}%
\else
\language=\csname l@#1\endcsname
\fi
#2}}
\providecommand{\BIBdecl}{\relax}
\BIBdecl

\bibitem{lientz1980software}
B.~P. Lientz and E.~B. Swanson, ``Software maintenance management,'' 1980.

\bibitem{de2005study}
S.~C.~B. de~Souza, N.~Anquetil, and K.~M. de~Oliveira, ``A study of the
  documentation essential to software maintenance,'' in \emph{Proceedings of
  the 23rd annual international conference on Design of communication:
  documenting \& designing for pervasive information}.\hskip 1em plus 0.5em
  minus 0.4em\relax ACM, 2005, pp. 68--75.

\bibitem{kajko2005survey}
M.~Kajko-Mattsson, ``A survey of documentation practice within corrective
  maintenance,'' \emph{Empirical Software Engineering}, vol.~10, no.~1, pp.
  31--55, 2005.

\bibitem{movshovitz2013natural}
D.~Movshovitz-Attias and W.~W. Cohen, ``Natural language models for predicting
  programming comments,'' 2013.

\bibitem{iyer2016summarizing}
S.~Iyer, I.~Konstas, A.~Cheung, and L.~Zettlemoyer, ``Summarizing source code
  using a neural attention model.'' in \emph{ACL (1)}, 2016.

\bibitem{yang2016query}
D.~Yang, A.~Hussain, and C.~V. Lopes, ``From query to usable code: An analysis
  of stack overflow code snippets,'' in \emph{Mining Software Repositories
  (MSR), 2016 IEEE/ACM 13th Working Conference on}.\hskip 1em plus 0.5em minus
  0.4em\relax IEEE, 2016, pp. 391--401.

\bibitem{nie2016query}
L.~Nie, H.~Jiang, Z.~Ren, Z.~Sun, and X.~Li, ``Query expansion based on crowd
  knowledge for code search,'' \emph{IEEE Transactions on Services Computing},
  vol.~9, no.~5, pp. 771--783, 2016.

\bibitem{nguyen2017automatic}
A.~T. Nguyen and T.~N. Nguyen, ``Automatic categorization with deep neural
  network for open-source java projects,'' in \emph{Proceedings of the 39th
  International Conference on Software Engineering Companion}.\hskip 1em plus
  0.5em minus 0.4em\relax IEEE Press, 2017, pp. 164--166.

\bibitem{oda2015learning}
Y.~Oda, H.~Fudaba, G.~Neubig, H.~Hata, S.~Sakti, T.~Toda, and S.~Nakamura,
  ``Learning to generate pseudo-code from source code using statistical machine
  translation (t),'' in \emph{Automated Software Engineering (ASE), 2015 30th
  IEEE/ACM International Conference on}.\hskip 1em plus 0.5em minus 0.4em\relax
  IEEE, 2015, pp. 574--584.

\bibitem{sridhara2010towards}
G.~Sridhara, E.~Hill, D.~Muppaneni, L.~Pollock, and K.~Vijay-Shanker, ``Towards
  automatically generating summary comments for java methods,'' in
  \emph{Proceedings of the IEEE/ACM international conference on Automated
  software engineering}.\hskip 1em plus 0.5em minus 0.4em\relax ACM, 2010, pp.
  43--52.

\bibitem{allamanis2016convolutional}
M.~Allamanis, H.~Peng, and C.~Sutton, ``A convolutional attention network for
  extreme summarization of source code,'' in \emph{International Conference on
  Machine Learning}, 2016, pp. 2091--2100.

\bibitem{haije2016automatic}
T.~Haije, B.~O.~K. Intelligentie, E.~Gavves, and H.~Heuer, ``Automatic comment
  generation using a neural translation model,'' 2016.

\bibitem{hochreiter1997long}
S.~Hochreiter and J.~Schmidhuber, ``Long short-term memory,'' \emph{Neural
  computation}, vol.~9, no.~8, pp. 1735--1780, 1997.

\bibitem{ranzato2015sequence}
M.~Ranzato, S.~Chopra, M.~Auli, and W.~Zaremba, ``Sequence level training with
  recurrent neural networks,'' \emph{arXiv preprint arXiv:1511.06732}, 2015.

\bibitem{baxter1998clone}
I.~D. Baxter, A.~Yahin, L.~Moura, M.~Sant'Anna, and L.~Bier, ``Clone detection
  using abstract syntax trees,'' in \emph{Software Maintenance, 1998.
  Proceedings., International Conference on}.\hskip 1em plus 0.5em minus
  0.4em\relax IEEE, 1998, pp. 368--377.

\bibitem{tai2015improved}
K.~S. Tai, R.~Socher, and C.~D. Manning, ``Improved semantic representations
  from tree-structured long short-term memory networks,'' \emph{arXiv preprint
  arXiv:1503.00075}, 2015.

\bibitem{rosenfeld2000two}
R.~Rosenfeld, ``Two decades of statistical language modeling: Where do we go
  from here?'' \emph{Proceedings of the IEEE}, vol.~88, no.~8, pp. 1270--1278,
  2000.

\bibitem{mnih2012fast}
A.~Mnih and Y.~W. Teh, ``A fast and simple algorithm for training neural
  probabilistic language models,'' \emph{arXiv preprint arXiv:1206.6426}, 2012.

\bibitem{li2015gated}
Y.~Li, D.~Tarlow, M.~Brockschmidt, and R.~Zemel, ``Gated graph sequence neural
  networks,'' \emph{arXiv preprint arXiv:1511.05493}, 2015.

\bibitem{bahdanau2014neural}
D.~Bahdanau, K.~Cho, and Y.~Bengio, ``Neural machine translation by jointly
  learning to align and translate,'' \emph{arXiv preprint arXiv:1409.0473},
  2014.

\bibitem{sutton1998introduction}
R.~S. Sutton and A.~G. Barto, \emph{Introduction to reinforcement
  learning}.\hskip 1em plus 0.5em minus 0.4em\relax MIT press Cambridge, 1998,
  vol. 135.

\bibitem{williams1992simple}
R.~J. Williams, ``Simple statistical gradient-following algorithms for
  connectionist reinforcement learning,'' in \emph{Reinforcement
  Learning}.\hskip 1em plus 0.5em minus 0.4em\relax Springer, 1992, pp. 5--32.

\bibitem{watkins1992q}
C.~J. Watkins and P.~Dayan, ``Q-learning,'' \emph{Machine learning}, vol.~8,
  no. 3-4, pp. 279--292, 1992.

\bibitem{keneshloo2018deep}
Y.~Keneshloo, T.~Shi, C.~K. Reddy, and N.~Ramakrishnan, ``Deep reinforcement
  learning for sequence to sequence models,'' \emph{arXiv preprint
  arXiv:1805.09461}, 2018.

\bibitem{konda2000actor}
V.~R. Konda and J.~N. Tsitsiklis, ``Actor-critic algorithms,'' in
  \emph{Advances in neural information processing systems}, 2000, pp.
  1008--1014.

\bibitem{silver2016mastering}
D.~Silver, A.~Huang, C.~J. Maddison, A.~Guez, L.~Sifre, G.~Van Den~Driessche,
  J.~Schrittwieser, I.~Antonoglou, V.~Panneershelvam, M.~Lanctot, and
  S.~Dieleman, ``Mastering the game of go with deep neural networks and tree
  search,'' \emph{Nature}, vol. 529, no. 7587, pp. 484--489, 2016.

\bibitem{aho1986compilers}
A.~V. Aho, R.~Sethi, and J.~D. Ullman, ``Compilers, principles, techniques,''
  \emph{Addison Wesley}, vol.~7, no.~8, p.~9, 1986.

\bibitem{weisupervised}
H.~H. Wei and M.~Li, ``Supervised deep features for software functional clone
  detection by exploiting lexical and syntactical information in source code,''
  2017.

\bibitem{papineni2002bleu}
K.~Papineni, S.~Roukos, T.~Ward, and W.~J. Zhu, ``Bleu: a method for automatic
  evaluation of machine translation,'' in \emph{Proceedings of the 40th annual
  meeting on association for computational linguistics}.\hskip 1em plus 0.5em
  minus 0.4em\relax Association for Computational Linguistics, 2002, pp.
  311--318.

\bibitem{schulman2015high}
J.~Schulman, P.~Moritz, S.~Levine, M.~Jordan, and P.~Abbeel, ``High-dimensional
  continuous control using generalized advantage estimation,'' \emph{arXiv
  preprint arXiv:1506.02438}, 2015.

\bibitem{duchi2011adaptive}
J.~Duchi, E.~Hazan, and Y.~Singer, ``Adaptive subgradient methods for online
  learning and stochastic optimization,'' \emph{Journal of Machine Learning
  Research}, vol.~12, no. Jul, pp. 2121--2159, 2011.

\bibitem{barone2017parallel}
A.~V.~M. Barone and R.~Sennrich, ``A parallel corpus of python functions and
  documentation strings for automated code documentation and code generation,''
  \emph{arXiv preprint arXiv:1707.02275}, 2017.

\bibitem{martin2009clean}
R.~C. Martin, \emph{Clean code: a handbook of agile software
  craftsmanship}.\hskip 1em plus 0.5em minus 0.4em\relax Pearson Education,
  2009.

\bibitem{banerjee2005meteor}
S.~Banerjee and A.~Lavie, ``Meteor: An automatic metric for mt evaluation with
  improved correlation with human judgments,'' in \emph{Proceedings of the acl
  workshop on intrinsic and extrinsic evaluation measures for machine
  translation and/or summarization}, vol.~29, 2005, pp. 65--72.

\bibitem{lin2004rouge}
C.~Y. Lin, ``Rouge: A package for automatic evaluation of summaries,''
  \emph{Text Summarization Branches Out}, 2004.

\bibitem{vedantam2015cider}
R.~Vedantam, C.~Lawrence~Zitnick, and D.~Parikh, ``Cider: Consensus-based image
  description evaluation,'' in \emph{Proceedings of the IEEE conference on
  computer vision and pattern recognition}, 2015, pp. 4566--4575.

\bibitem{sutskever2014sequence}
I.~Sutskever, O.~Vinyals, and Q.~V. Le, ``Sequence to sequence learning with
  neural networks,'' in \emph{Advances in neural information processing
  systems}, 2014, pp. 3104--3112.

\bibitem{eriguchi2016tree}
A.~Eriguchi, K.~Hashimoto, and Y.~Tsuruoka, ``Tree-to-sequence attentional
  neural machine translation,'' \emph{arXiv preprint arXiv:1603.06075}, 2016.

\bibitem{kilickaya2016re}
M.~Kilickaya, A.~Erdem, N.~Ikizler-Cinbis, and E.~Erdem, ``Re-evaluating
  automatic metrics for image captioning,'' \emph{arXiv preprint
  arXiv:1612.07600}, 2016.

\bibitem{gu2016deep}
X.~Gu, H.~Zhang, D.~Zhang, and S.~Kim, ``Deep api learning,'' in
  \emph{Proceedings of the 2016 24th ACM SIGSOFT International Symposium on
  Foundations of Software Engineering}.\hskip 1em plus 0.5em minus 0.4em\relax
  ACM, 2016, pp. 631--642.

\bibitem{mou2016convolutional}
L.~Mou, G.~Li, L.~Zhang, T.~Wang, and Z.~Jin, ``Convolutional neural networks
  over tree structures for programming language processing.'' in \emph{AAAI},
  vol.~2, no.~3, 2016, p.~4.

\bibitem{piech2015learning}
C.~Piech, J.~Huang, A.~Nguyen, M.~Phulsuksombati, M.~Sahami, and L.~Guibas,
  ``Learning program embeddings to propagate feedback on student code,''
  \emph{arXiv preprint arXiv:1505.05969}, 2015.

\bibitem{parisotto2016neuro}
E.~Parisotto, A.~Mohamed, R.~Singh, L.~Li, D.~Zhou, and P.~Kohli,
  ``Neuro-symbolic program synthesis,'' \emph{arXiv preprint arXiv:1611.01855},
  2016.

\bibitem{maddison2014structured}
C.~Maddison and D.~Tarlow, ``Structured generative models of natural source
  code,'' in \emph{International Conference on Machine Learning}, 2014, pp.
  649--657.

\bibitem{dam2016deep}
H.~K. Dam, T.~Tran, and T.~Pham, ``A deep language model for software code,''
  \emph{arXiv preprint arXiv:1608.02715}, 2016.

\bibitem{ling2016latent}
W.~Ling, E.~Grefenstette, K.~M. Hermann, T.~Ko{\v{c}}isk{\`y}, A.~Senior,
  F.~Wang, and P.~Blunsom, ``Latent predictor networks for code generation,''
  \emph{arXiv preprint arXiv:1603.06744}, 2016.

\bibitem{allamanis2015bimodal}
M.~Allamanis, D.~Tarlow, A.~Gordon, and Y.~Wei, ``Bimodal modelling of source
  code and natural language,'' in \emph{International Conference on Machine
  Learning}, 2015, pp. 2123--2132.

\bibitem{sutton2000policy}
R.~S. Sutton, D.~A. McAllester, S.~P. Singh, and Y.~Mansour, ``Policy gradient
  methods for reinforcement learning with function approximation,'' in
  \emph{Advances in neural information processing systems}, 2000, pp.
  1057--1063.

\bibitem{mnih2015human}
V.~Mnih, K.~Kavukcuoglu, D.~Silver, A.~A. Rusu, J.~Veness, M.~G. Bellemare,
  A.~Graves, M.~Riedmiller, A.~K. Fidjeland, G.~Ostrovski \emph{et~al.},
  ``Human-level control through deep reinforcement learning,'' \emph{Nature},
  vol. 518, no. 7540, pp. 529--533, 2015.

\bibitem{zhu2017target}
Y.~Zhu, R.~Mottaghi, E.~Kolve, J.~J. Lim, A.~Gupta, L.~Fei-Fei, and A.~Farhadi,
  ``Target-driven visual navigation in indoor scenes using deep reinforcement
  learning,'' in \emph{Robotics and Automation (ICRA), 2017 IEEE International
  Conference on}.\hskip 1em plus 0.5em minus 0.4em\relax IEEE, 2017, pp.
  3357--3364.

\bibitem{ren2017deep}
Z.~Ren, X.~Wang, N.~Zhang, X.~Lv, and L.~J. Li, ``Deep reinforcement
  learning-based image captioning with embedding reward,'' in \emph{Computer
  Vision and Pattern Recognition (CVPR), 2017 IEEE Conference on}.\hskip 1em
  plus 0.5em minus 0.4em\relax IEEE, 2017, pp. 1151--1159.

\bibitem{li2016deep}
J.~Li, W.~Monroe, A.~Ritter, M.~Galley, J.~Gao, and D.~Jurafsky, ``Deep
  reinforcement learning for dialogue generation,'' \emph{arXiv preprint
  arXiv:1606.01541}, 2016.

\bibitem{zhang2017sentence}
X.~Zhang and M.~Lapata, ``Sentence simplification with deep reinforcement
  learning,'' in \emph{Proceedings of the 2017 Conference on Empirical Methods
  in Natural Language Processing}, 2017, pp. 584--594.

\end{thebibliography}
\end{document}